
\input psfig
\input mn.tex

\def\xG{x_{\rm G}}
\def\sigmaG{\sigma_{\rm G}}

\def\piby2{{\pi \over 2}}

\def\vcirc{v_{\rm circ}}

\def\sech{\rm sech}
\def\csch{\rm csch}

\def\as{a_{\rm s}}
\def\rhos{\rho_{\rm s}}
\def\rhos{\rho_{\rm s}}
\def\rhat{{\bf {\hat r}}}
\def\shat{{\bf {\hat s}}}

\def\sqarcmin{\,{\rm arcmin}^2}
\def\Mmin{M_{\rm min}}
\def\Mmax{M_{\rm max}}
\def\smin{s_{\rm min}}
\def\smax{s_{\rm max}}
\def\as{a_{\rm s}}
\def\mt{m_{\rm t}}
\def\logten{\log_{10}}
\def\muay{{\mu}{\rm as\,yr}^{-1}}

\def\kms{km\,s$^{-1}$}
%
%
\def\spose#1{\hbox to 0pt{#1\hss}}
\def\lta{\mathrel{\spose{\lower 3pt\hbox{$\sim$}}
    \raise 2.0pt\hbox{$<$}}}
\def\gta{\mathrel{\spose{\lower 3pt\hbox{$\sim$}}
    \raise 2.0pt\hbox{$>$}}}
\def\today{\ifcase\month\or
 January\or February\or March\or April\or May\or June\or
 July\or August\or September\or October\or November\or December\fi
 \space\number\day, \number\year}
\newdimen\hssize
\hssize=8.4truecm  
\newdimen\hdsize
\hdsize=17.7truecm    


\newcount\eqnumber
\eqnumber=1
\def\chaphead{}
 
\def\new{\hbox{(\rm\chaphead\the\eqnumber)}\global\advance\eqnumber by 1}
 
\def\first{\hbox{(\rm\chaphead\the\eqnumber a)}\global\advance\eqnumber by 1}
\def\last#1{\advance\eqnumber by -1 \hbox{(\rm\chaphead\the\eqnumber#1)}\advance
     \eqnumber by 1}
 
\def\ref#1{\advance\eqnumber by -#1 \chaphead\the\eqnumber
     \advance\eqnumber by #1}
 
\def\nref#1{\advance\eqnumber by -#1 \chaphead\the\eqnumber
     \advance\eqnumber by #1}

\def\eqnam#1{\xdef#1{\chaphead\the\eqnumber}}
 
 

\pageoffset{-0.85truecm}{-1.05truecm}



\pagerange{}
\pubyear{version: \today}
\volume{}


\begintopmatter

\title{The Present and Future Mass of the Milky Way Halo}

\author{M.I.\ Wilkinson and N.W.\ Evans}

\vskip0.15truecm
\affiliation{Theoretical Physics, Department of Physics, 1 Keble Road,
                 Oxford, OX1 3NP}

\shortauthor{M.I.\ Wilkinson and N.W.\ Evans} 

\shorttitle{The Present and Future Mass of the Milky Way Halo}



\abstract{A simple model for the Milky Way halo is presented. It
has a flat rotation curve in the inner regions, but the density falls
off sharply beyond an outer edge. This truncated, flat rotation curve
(TF) model possesses a rich family of simple distribution functions
which vary in velocity anisotropy.  The model is used to estimate the
total mass of the Milky Way halo using the latest data on the motions
of satellite galaxies and globular clusters at Galactocentric radii
greater than 20 kpc. This comprises a dataset of 27 objects with known
distances and radial velocities, of which 6 also possess measured
proper motions.  Unlike earlier investigations, {\it we find entirely
consistent maximum likelihood solutions unaffected by the presence or
absence of Leo~I, provided both radial and proper motion data are
used.} The availability of the proper motion data for the satellites
is crucial as, without them, the mass estimates with and without Leo~I
are inconsistent at the $99\%$ confidence level. All these results are
derived from models in which the velocity normalisation of the halo
potential is taken as $\sim 220$ \kms.

A detailed analysis of the uncertainties in our estimate is presented,
including the effects of the small dataset, possible incompleteness or
correlations in the satellite galaxy sample and the measurement
errors. The most serious uncertainties come from the size of the
dataset, which may cause a systematic underestimate by a factor of
two, and the measurement errors, which cause a scatter in the mass of
the order of a factor of two.  We conclude that the total mass of the
halo is $\sim 1.9^{+3.6}_{-1.7} \times 10^{12} M_{\odot}$, while the
mass within 50 kpc is $\sim 5.4^{+0.2}_{-3.6} \times 10^{11}
M_{\odot}$. In the near future, ground-based radial velocity surveys
of samples of blue horizontal branch (BHB) stars are a valuable way to
augment the sparse dataset. A dataset of $\sim 200$ radial velocities
of BHB stars will reduce the uncertainty in the mass estimate to $\sim
20\%$. In the coming decade, microarcsecond astrometry will be
possible with the {\it Space Interferometry Mission} (SIM) and the
{\it Global Astrometry Interferometer for Astrophysics} (GAIA)
satellites. For example, GAIA can provide the proper motions of the
the distant dwarfs like Leo~I to within $\pm 15$ \kms and the nearer
dwarfs like Ursa Minor to within $\pm 1$ \kms. This will also allow
the total mass of the Milky Way to be found to $\sim 20\%$. SIM and
GAIA will also provide an accurate estimate of the velocity
normalisation of the halo potential at large radii.}


\maketitle  

%
\section{Introduction}
The aim of this paper is to obtain a consistent estimate of the total
mass of the Milky Way halo. The structure and extent of the dark
matter halos of galaxies is a matter of great strategic importance for
modern astrophysics.  Of course, it is especially important to extract
as much information as possible about the halo of our Galaxy, the
proximity of which allows it to be studied in exceptional detail.
Unfortunately, the mass and size of the Milky Way halo are amongst the
most poorly known of all Galactic parameters. They are much more
uncertain than the distance to the Galactic Centre or the Oort's
constants, for example.

The Milky Way's gas rotation curve cannot be traced beyond $\sim 20$
kpc, and so it is natural to look to the kinematics of stellar tracers
of the distant halo for estimates of the mass.  The motions of the
bound satellites of the Milky Way, together with the globular
clusters, contain valuable information about the halo potential in
which they are moving. Given a model of the gravity field, it is
possible to constrain the values of parameters such as the halo's
extent, total mass and velocity anisotropy using the radial velocities
and proper motions of the distant satellites and globular clusters. A
number of authors have studied this problem (e.g., Little \& Tremaine
1987; Zaritsky et al. 1989; Kulessa \& Lynden-Bell 1992; Kochanek
1996), obtaining a variety of different mass estimates. One
peculiarity of all previous studies, however, is the sensitivity of
the mass estimates to whether or not Leo~I is bound to the Milky
Way. Leo~I is unusual in that it has one of the largest radial
velocities despite being the second most distant of the Milky Way
satellites. It is evidently desirable to obtain mass estimates which
do not depend strongly on the velocity of a single satellite.

The current dataset of satellites and globular clusters at distances
greater than 20 kpc from the Galactic Centre contains only $27$
objects, of which just 6 have measured proper motions. It has been
argued that the dataset on the satellite galaxies is complete (Pryor
1998), though this is perhaps unclear as undiscovered satellites could
still be lurking within the Zone of Avoidance, especially at large
radii. Even so, simple scaling arguments applied to the volume of the
Zone of Avoidance suggest that the number of undiscovered satellites
within $\sim 250$ kpc is $\lta 5$.  The dataset has changed in recent
years only through painstaking measurements of the proper motions of
some of the closest satellites, such as Ursa Minor and Draco (Scholz
\& Irwin 1994; Schweitzer, Cudworth \& Majewski 1997). However, as there
are now real hopes that the next few years will see some substantial
progress, it is timely to re-examine the problem in order to determine
the avenues along which most progress is liable to be made. First, the
{\it Space Interferometry Mission} or SIM (Unwin, Boden, Shao 1997)
offers the possibility of microarcsecond parallaxes and proper
motions, albeit with long integration times for objects as faint as
the dwarf satellites.  Second, the {\it Global Astrometry
Interferometer for Astrophysics} or GAIA (Lindegren \& Perryman 1996)
will be able find the space motions of all the satellite galaxies to
within $10 \%$, though the distances will be less certain. Third,
large samples of blue horizontal branch stars that contaminate quasar
surveys are becoming available (Flynn, Sommer-Larsen, Christensen \&
Hawkins 1995; Warren 1998, private communication; Miller 1998, private
communication). These are much more numerous than the satellite
galaxies and -- even though just their radial velocities and distances
will be available -- they are an invaluable addition to the dataset.

In Section 2, we present our model of the Milky Way halo and describe
its properties in some detail.  Our strategy for finding the total
mass of the Milky Way halo using the radial velocity and proper motion
data for the satellite galaxies and distant globular clusters is also
reported.  In Sections 3 and 4, we present the results of our analysis
of the present dataset. With such a limited amount of data, we must be
cautious in interpreting the results of statistical analyses.  In
order to estimate the amount of uncertainty present in mass estimates,
we generate large numbers of synthetic datasets and analyse them in
the same way as the real data.  Section 5 identifies three main
sources of uncertainty -- measurement errors, modelling uncertainties
and correlations in the dataset -- and examines the effect of each in
turn to place an error estimate on the mass of the Milky Way.
Finally, Section 6 turns to the future and evaluates the best
strategies to exploit the expected new information from astrometric
satellites and ground-based radial velocity surveys of halo stars.
\beginfigure{1}
\centerline{\psfig{figure=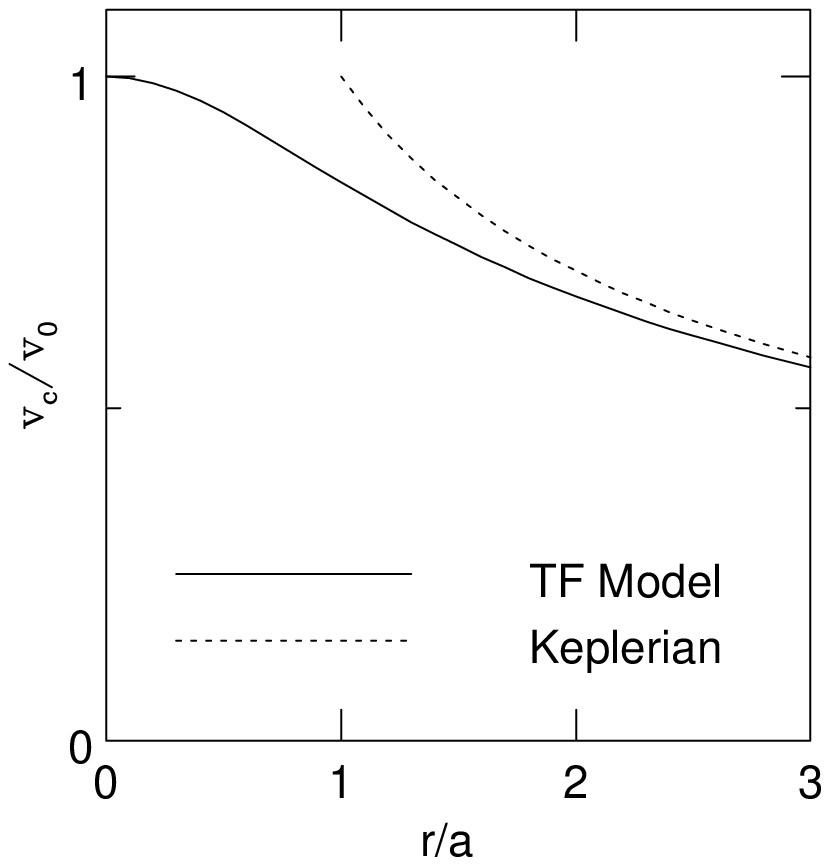,width=\hssize}}
\smallskip\noindent
\caption{{\bf Figure 1.} Rotation curve of the TF model (solid
line) compared with a Keplerian rotation curve (broken line).  As $r
\rightarrow \infty$, the rotation curve of the TF model approaches the
Keplerian one.}
\endfigure
\section{The Algorithm}
This section is mainly theoretical and discusses in turn our model for
the Milky Way halo in Section 2.1 and the tracer population of satellites
in Section 2.2. The maximum likelihood algorithm for the mass of the
halo is reported in Section 2.3.

\subsection{The Truncated, Flat Rotation Curve (TF) Model}
In order to study the dynamical properties of a halo model, it is
necessary to know the phase space distribution function (DF). This
depends only on the isolating integrals of motion via Jeans (1919)
theorem. The isotropic DF depends only on the binding energy per unit
mass $\varepsilon$. If $\rho(r)$ is the density of the model and
$\psi(r)$ is the corresponding potential, then the isotropic
distribution function is given by the well-known formula (Eddington
1915; Binney \& Tremaine 1987)
\eqnam{\isoDF}
$$F(\varepsilon) = {1 \over \sqrt{8}\pi^2}{d \over d\varepsilon}
                 \int_0^\varepsilon
{d\rho \over d\psi}{d\psi \over \sqrt{\varepsilon - \psi}}.\eqno\new$$
This is the simplest possible case, but we can also look for
anisotropic DFs which depend on the angular momentum per unit mass
$l$. A particularly simple and attractive Ansatz is (e.g., 
H\'enon 1973; Dejonghe 1986)
\eqnam{\dejonghe}
$$F(\varepsilon,l) = l^{-2 \beta} f(\varepsilon),\eqno\new$$
where
\eqnam{\anisoDF}
$$\eqalign{f(\varepsilon) = & {2^{\beta - 3/2} \over \pi^{3/2} 
\Gamma [m - 1/2 + \beta] \Gamma [1 - \beta]} 
{d \over d\varepsilon}\cr  & \times 
\int_o^\varepsilon d \psi {d^m r^{2\beta} \rho \over d \psi^m} 
 (\varepsilon - \psi)^{\beta - 3/2 + m}.}\eqno\new$$ 
Here, $m$ is an integer whose value is chosen such that the integral
in (\anisoDF) converges. For such a DF, the velocity dispersions
$\langle v^2_{\phi}\rangle$ and $\langle v^2_{\theta}\rangle$ are
equal, and there is a constant orbital anisotropy $\beta = 1 -
\langle v_{\theta}^2\rangle/\langle v_r^2 \rangle$. 

Clearly, the construction of the DF is very much simpler if the
density $\rho$ can be written as an explicit function of the potential
$\psi$.  There are few such simple models known -- although famous
ones have been discovered by Jaffe (1983), Hernquist (1990), Evans
(1994) and Zhao (1996). We now present another example. This model has
a flat rotation curve in the inner regions and at large radii, the
density falls off abruptly like $r^{-5}$. For this reason, we shall
call it {\it the truncated, flat rotation curve model}, henceforth TF.

The density of the TF model is
\eqnam{\density1}
$$\rho(r) = {M\over 4\pi}{a^2\over r^2 (r^2 + a^2)^{3/2}}.\eqno\new$$
and the potential, which can easily be obtained from Poisson's equation, is
\eqnam{\potential1}
$$\psi(r) = {GM\over a}\log\Bigl({\sqrt{r^2 + a^2} + a \over r}\Bigr).
\eqno\new$$
This model is similar to Jaffe's in that the density is cusped like
$r^{-2}$ in the nucleus; it differs in that the density falls off
like $r^{-5}$ rather than like $r^{-4}$ in the outer reaches.  The
rotation curve is flat with amplitude $v_0 = \sqrt{GM/a}$ in the inner
parts. The general rule is
\eqnam{\vzero}
$$v_{\rm circ}^2 = {v_0^2\over ( 1 + r^2/a^2)^{1/2}}.\eqno\new$$
As Fig.~1 illustrates, the rotation curve becomes Keplerian for
$r>>a$.  The density can be written in terms of the dimensionless
potential $\phi = \psi/v_0^2$ as
$$\rho(\phi) = {M\over 4\pi a^3} {\sinh^5 \phi \over
\cosh^3 \phi}.\eqno\new$$
This follows because (\potential1) can be inverted as
\eqnam{\crux}
$$r(\phi) = a \csch \phi.\eqno\new$$
which is the crucial equation on which the value of the model rests.
So, the isotropic DF from equation (\isoDF) is
$$\eqalign{F(\varepsilon) = {M \over 2\sqrt{2} \pi^3 a^3 v_0^3}
\int_0^\varepsilon & {d\phi \over (\varepsilon-\phi)^{1/2}} \{ \sinh^2
\phi \tanh \phi \cr & +
\tanh^3 \phi + 3 \tanh^3 \phi \sech^2 \phi \}.}\eqno\new$$
This is not a tractable integral, but the asymptotic behaviour is
easily derived. The approximate form of the DF in the envelope
($\varepsilon \rightarrow 0$) is $F(\varepsilon) \sim
\varepsilon^{7/2}$. Near the central cusp ($\varepsilon 
\rightarrow \infty$), the DF becomes $F(\varepsilon) \sim 
\exp(2\varepsilon)$. 
\beginfigure{2}
\centerline{\psfig{figure=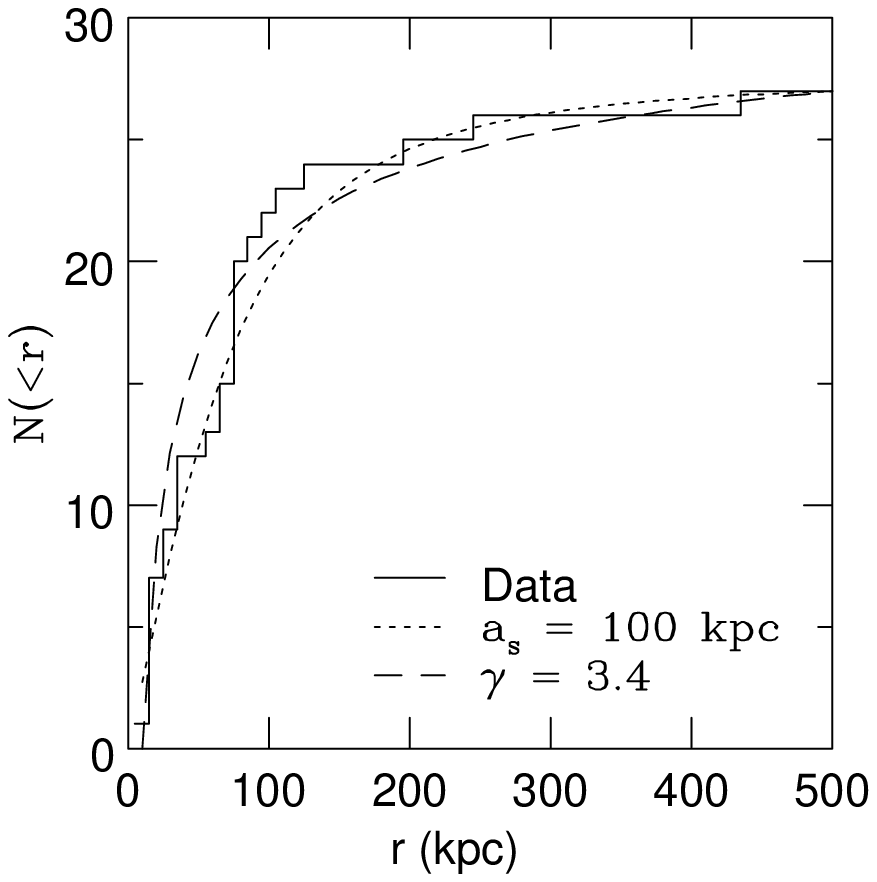,height=\hsize}}
\smallskip\noindent
\caption{{\bf Figure 2.} The cumulative number $N(<r)$ of the
satellites and distant globular clusters is plotted against
Galactocentric radius $r$. Superposed are the best fitting shadow
(dotted line) and power-law (dashed line) tracer models.  }
\endfigure
\beginfigure{3}
\centerline{\psfig{figure=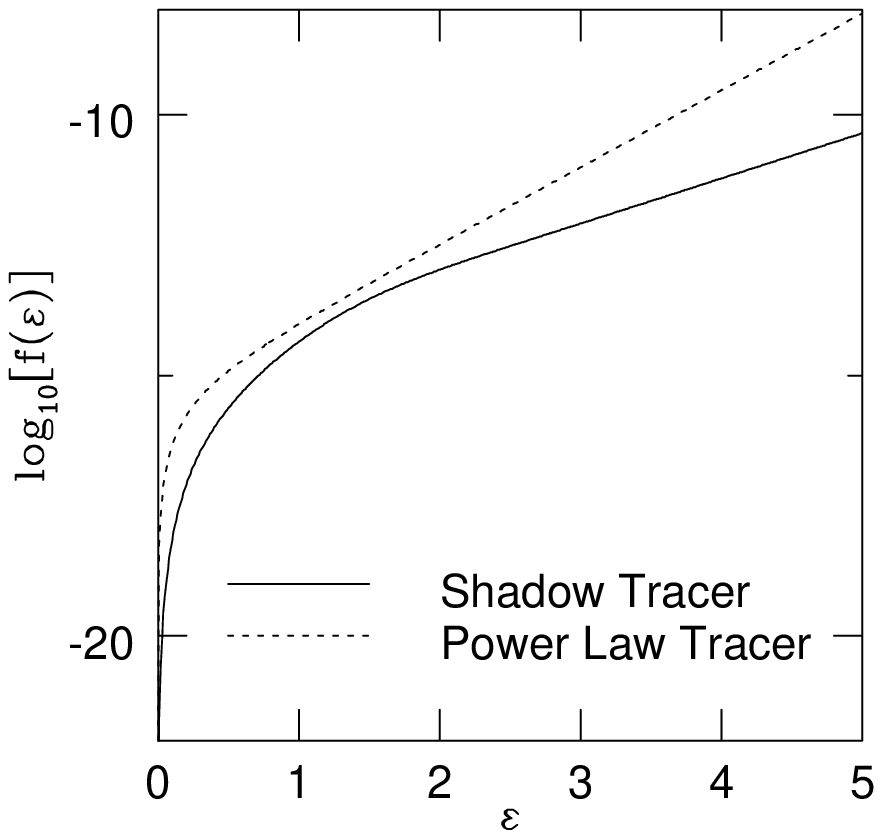,width=\hssize}}
\smallskip\noindent
\caption{{\bf Figure 3.} Logarithm of the isotropic DF for a
shadow tracer model (solid line) compared with that for a power-law
tracer model (broken line).}
\endfigure
\begintable*{1}
\caption{{\bf Table 1.} Probability formulae (15) and (16) are given
explicitly for the shadow tracer and power-law tracer
populations. In each case, $P(r_i,v_{ri}|a,\beta)$ is the
probability when only a radial velocity $v_{ri}$ is available and
$P(r_i,v_{i}|a,\beta)$ is the corresponding probability when the full
space velocity $v_i$ is available from proper motion
measurements.}
\halign{#\hfil&\quad#\hfil\cr
\noalign{\hrule}
\noalign{\vskip0.1truecm}
Satellite Model & \hskip6truecm Probability\cr 
\noalign{\vskip0.1truecm}
Shadow Tracer & $P(r_i,v_{ri}|a,\beta) = {\as^2a^{2\beta-2} \over
                4\sqrt{2}\pi^2\rhos r^{2\beta}v_0}
                \int_0^{\varepsilon_m}d\phi  
		{[(5-2\beta)a^2 - (2\beta-2)\as^2\sinh^2 \phi]\cosh\phi
                \over (\varepsilon_m - \phi)^{1/2} 
                \sinh^{2\beta-4}\phi\,
		(a^2+ \as^2\sinh^2 \phi)^{5/2}}$\cr 
\noalign{\vskip0.1truecm}
	      &	$P(r_i,v_{i}|a,\beta) = {l^{-2\beta} \over \rhos}
		{2^{\beta -5/2} 
		v_0^{2\beta -3}a^{2\beta} \over
		\pi^{5/2} \Gamma[3/2+\beta]\Gamma[1-\beta]}
		\int_0^{{\pi/2}} d\theta\, \sin\theta\cos\theta 
		{d \over d\varepsilon}\Bigl[\varepsilon \Bigl( 
		{d^2 \over d\phi^2}
	\Bigl[{\sinh^{5-2\beta}\phi\over(a^2 + \as^2\sinh^2 \phi)^{3/2}}
		\Bigr](\varepsilon-\phi)^{\beta+1/2}\Bigr)_{\phi \rightarrow
		\varepsilon\sin^2\theta} \Bigr]$\cr
\noalign{\vskip0.2truecm}
Power-Law Tracer & $P(r_i,v_{ri}|a,\beta) = 
		   {a^{2\beta-\gamma} (\gamma-2\beta)\over \sqrt{2}\pi
		   r^{2\beta-\gamma} v_0}
		   \int_0^{\varepsilon_m} d\phi {\sinh^{\gamma - 2\beta-1}\phi
		   \cosh\phi \over (\varepsilon_m - \phi)^{1/2}}$\cr
\noalign{\vskip0.1truecm}
		 & $P(r_i,v_{i}|a,\beta) = 
		   {l^{-2\beta} r^\gamma 2^{\beta - 1/2}
		   v_0^{2\beta-3}a^{2\beta-\gamma}\over \pi^{3/2} 
		   \Gamma[{3/2}+\beta]\Gamma[1-\beta]}
		   \int_0^{{\pi/2}} d\theta \sin\theta
		   \cos\theta {d \over d\varepsilon}
		   \Bigl[\varepsilon \Bigl({d^2 \over
		   d\phi^2} \Bigl[ \sinh^{\gamma -2\beta}\phi \Bigr] 
		   (\varepsilon-\phi)^{\beta+1/2}\Bigr)_{\phi 
		   \rightarrow
		   \varepsilon\sin^2\theta}\Bigr]$\cr
\noalign{\vskip0.1truecm}\noalign{\hrule}
}
\endtable
The velocity dispersions of the isotropic model are
$$\eqalign{\langle v_r^2 \rangle  = \langle v_\theta^2 \rangle =
\langle v_\phi^2
\rangle & = {v_0^2 (r^2 + a^2)^{1/2} \over 2 a^4}\Bigl[a (2 r^2 +
a^2)\cr  & 
- {2r^2 \over a} (r^2 + a^2) \log[{r^2+a^2 \over r^2}]\Bigr].}\eqno\new$$
A careful Taylor expansion shows that as $r \rightarrow 0$, $\langle
v_r^2 \rangle \rightarrow v_0^2/2$.  The same property is possessed by
the Jaffe sphere. This is because the central parts of both models are
very similar and resemble the singular isothermal sphere. As $r
\rightarrow \infty$, $\langle v_r^2 \rangle \rightarrow 0$, as
expected.  The circular orbit model has $\langle v_r^2 \rangle =0$ and
$\langle v_\theta^2 \rangle = \langle v_\phi^2
\rangle = {1\over 2} \vcirc^2$. The radial velocity dispersion for a
model with constant anisotropy $\beta = 1 - \langle v_{\theta}^2
\rangle / \langle v_r^2 \rangle$ is
$$\langle v_r^2 \rangle = {M a^{2\beta-3} v_0^2 \over 4 \pi r^{2\beta}
\rho} \int_0^{\phi(r)} {\sinh^{5-2\beta} \phi \over \cosh^3 \phi}
d\phi.\eqno\new$$
The TF potential has been used before by Lin and Lynden-Bell (1982) in
their work on the orbits of the Magellanic clouds and by Lynden-Bell
\& Lynden-Bell (1995) in their study of the kinematics of halo
streams, but its DF does not appear to be available in the
literature. In many respects, the TF model rivals the Jaffe sphere in
terms of simplicity and usefulness.
\subsection{The Satellite Number Density}
Let us assume that the Milky Way halo is a TF model whose total mass
$M$ is to be found. For the analysis of the distant satellites, the
most important thing is not the DF of the self-consistent mass density
(\density1) but the DF of a tracer population. For this, we consider
two distinct possibilities. 

First, the density of the satellites may ``shadow'' the total density of
the halo. In the case of such {\it shadow tracers}, the number density
of the satellites is given by
\eqnam{\density1}
$$\rhos(r) \propto {\as^2\over r^2 (r^2 + \as^2)^{3/2}}.\eqno\new$$
In other words, the satellites follow another TF model with a
scalelength $\as$ which may or may not be the same as that of the
halo. The cumulative number of satellites $N(<r)$ within radius $r$
is plotted in Fig.~2.  Here, the data are the 27 satellite galaxies
and globular clusters at Galactocentric radii greater than 20
kpc. Superposed is the best fitting shadow tracer in dotted lines, for
which the scalelength $\as$ is 100 kpc.  The second alternative is
that the number density of the satellites is a scale-free power-law,
i.e.,
$$\rhos(r)  \propto {1 \over r^\gamma},\eqno\new$$
where $\gamma$ is the asymptotic density fall-off. We assume that this
law holds good beyond a lower cut-off (to evade the singularity at the
origin) and sometimes even an outer cut-off.  We shall call this {\it
the power-law tracer} case.  The best fitting power-law tracer is also
shown in Fig.~2. It has $\gamma = 3.4$ so that the density indeed
falls off like a typical spheroid population.

For the shadow and power-law tracers, the isotropic DF is plotted as a
function of binding energy per unit mass in Fig.~3. Anticipating the
results in the next Section, the halo is assumed to be a TF model with
scalelength $a = 240$ kpc for the shadow tracer case and $a=170$ kpc
for the power-law tracer case. In the figure, the shadow tracer
population is a TF model with $a = 100$ kpc and the power-law tracer
model has $\gamma = 3.4$. More general families of anisotropic DFs
(\dejonghe) are also available as simple quadratures by expressing
$\rhos(r)$ in terms of the potential using (\crux) and substituting
into (\anisoDF). We shall not give the formulae here, but proceed to
construct the needed probabilities directly.

\subsection{The Bayesian Likelihood Method}
The models contain at least two free parameters, namely $\beta$ which
fixes the anisotropy of the orbits and $M$ which is the total mass of
the Milky Way halo. This section outlines our strategy for
constraining the model parameters using the radial velocities and
proper motions of the Milky Way satellites. The method was proposed by
Little \& Tremaine (1987) and developed further by Kochanek (1996). Of
course, the mass $M$ depends on the scalelength $a$ through
eq.~(\vzero), and in what follows all probabilities are quoted in
terms of $\beta$ and $a$.

Suppose for each of $N$ satellites at positions $r_i$ ($i = 1\dots N$)
we measure the radial velocity $v_{ri}$. Given a particular choice of
model parameters ($a$, $\beta$), the probability of finding a
satellite at radius $r_i$ moving with radial velocity $v_{ri}$ is
simply
\eqnam{\simpleprob}
$$P(r_i,v_{ri}|a,\beta) = {1 \over \rhos}\int d^3v\, l^{-2\beta}
f(\varepsilon)\delta(v_r - v_{ri}).\eqno\new$$
where $\rhos$ is the density distribution of the satellites.  Using
(\anisoDF) for $f(\varepsilon)$, it can be shown via Laplace
transforms that (e.g., Kochanek 1996)
\eqnam{\dataprob}
$$P(r_i,v_{ri}|a,\beta) = {1\over \sqrt{2}\pi\rhos
r^{2\beta}}\int_0^{\varepsilon_m}{d\psi \over (\varepsilon_m - \psi)^{1/2}}
{dr^{2\beta}\rhos \over d\psi} ,\eqno\new$$
if $\varepsilon_m = \psi - v_{ri}^2/2 > 0$ and zero otherwise. Note
that this expression holds for all values of $m$ in equation
(\anisoDF). If we also have the proper motion of a satellite, then we
can calculate its total velocity, $v_i$ and hence its tangential
velocity, $v_{ti} = v_i^2 - v_{ri}^2$. In this case the delta function
in equation (\simpleprob) becomes $\delta^3(\bf{v} - \bf{v_i})$ and
the probability is simply
\eqnam{\properprob}
$$P(r_i,v_{i}|a,\beta) = {f(\varepsilon)l^{-2\beta}\over
\rhos}.\eqno\new$$
if $\varepsilon = \psi - (v_{ri}^2 + v_{ti}^2)/2 > 0$ and zero
otherwise. Table~1 gives the expressions for the probabilities
(\dataprob) and (\properprob) for each of the two tracer populations.

\begintable{2}
\def\tableunderline{\noalign{\vskip0.1truecm}\noalign{\vskip0.1truecm}}
\caption{{\bf Table 2.} Data on the radial velocities of the
satellites and distant globular clusters. The sources are: $^{a}$
Harris (1996), $^{b}$ Mateo (1998). Listed are Galactic coordinates
($\ell,b$), heliocentric and Galactocentric distances ($s$ and
$r$) in kpc, heliocentric and Galactocentric line of sight radial
velocities ($v_\odot$ and $v_r$) in \kms, together with object
type.}
\halign{\hfil#\hfil&\quad\hfil#\hfil&\quad\hfil#\hfil&\quad\hfil#\hfil&\quad\hfil#\hfil&\quad\hfil#\hfil&\quad\hfil#\hfil&\quad\hfil#\hfil\cr
\noalign{\hrule}
\noalign{\vskip0.1truecm}
Name&$\ell$&    $b$&    $s$&    $r$ & $v_{\odot}$&  v$_r$& Type\cr
\tableunderline
Pal 13$^{a}$&    87&  -43&     26&             27&        -28&    138&   GC\cr
\tableunderline
NGC 5634$^{a}$&  342&  49&     25&             21&        -45&    -80&   GC\cr
\tableunderline
NGC 5824$^{a}$&  333&  22&     31&             25&        -38&    -127&  GC\cr
\tableunderline
NGC 5694$^{a}$&  331&  30&     34&             28&        -146&   -232& GC\cr
\tableunderline
NGC 6229$^{a}$&  74&   40&     29&             29&        -154&   22&   GC\cr
\tableunderline
Pal 15$^{a}$&    19&   24&     44&             37&        69&     148&  GC\cr
\tableunderline
NGC 7006$^{a}$&  64&   -19&    41&             38&        -378&   -180& GC\cr
\tableunderline
Pal 14$^{a}$&    29&   42&     72&             67&        77&     170&  GC\cr
\tableunderline
Eridanus$^{a}$&  218&  -41&    81&             86&        -24&    -141& GC\cr
\tableunderline
NGC 2419$^{a}$&  180&  25&     82&             90&        -20&    -27&  GC\cr
\tableunderline
Pal 4$^{a}$&     202&  72&     100&            102&       75&     51&   GC\cr
\tableunderline
AM-1$^{a}$&      258&  -48&    119&            120&       116&    -41&  GC\cr
\tableunderline
Pal 2$^{a}$&     171&  -9&     27&             35&       -133&
-105& GC\cr
\tableunderline
Arp 2$^{a}$&     9&     -21&   28&            20&        115&     153&
GC\cr
\tableunderline
NGC 7492$^{a}$&  53&    -63&   25&            24&        -208&
-128&  GC\cr
\tableunderline
Fornax$^{b}$&    237&   -66&   138&          140&       53&     -36&  dSph\cr
\tableunderline
Leo~I$^{b}$&     226&   49&     250&         254&       286&    178&  dSph\cr
\tableunderline
Leo~II$^{b}$&    220&   67&     205&         208&       76&     22&   dSph\cr
\tableunderline
Sextans$^{b}$&   244&   42&     86&           89&        227&    75&   dSph\cr
\tableunderline
Phoenix$^{b}$&   272&   -69&     445&         445&       56&     -34&  dIrr/dSph\cr
\tableunderline
Carina$^{b}$&    260&   -22&     101&         103&        224&    8&   dSph\cr
\tableunderline
\noalign{\vskip0.1truecm}\noalign{\hrule}
}
\endtable

\begintable*{3}
\def\tableunderline{\noalign{\vskip0.1truecm}\noalign{\vskip0.1truecm}}
\caption{{\bf Table 3.} Data on the radial velocities and proper
motions of the six satellites for which proper motions are
available. All proper motions are quoted as the heliocentric motions
$\mu_{\delta}$ and $\mu_{\alpha}\cos\delta$, where $\alpha$ and
$\delta$ and the R.A. and Dec., respectively. All proper motions are
in arcsec per century. $v_r$ and $v_t$ are the Galactocentric radial
and tangential velocities in \kms, calculated assuming
R$_{\odot}$ = 8.0 kpc and the motion of the sun to be (-9, 232, 11)
relative to the rest frame of the Galaxy.  Sources: 1 - Kroupa \&
Bastian (1997); 2 - Schweitzer, Cudworth, Majewski, \& Suntzeff
(1995); 3 - Schweitzer, Cudworth \& Majewski (1997); 4 - Odenkirchen,
Brosche, Geffert \& Tucholke (1997); 5 - Dauphole, Geffert, Colin,
Ducourant, Odenkirchen \& Tucholke (1996); 6 - Scholz \& Irwin (1994);
7 - Mateo (1998). Notes: 1 - The LMC/SMC have been treated as a single
object moving with the motion of the LMC but located at the centre of
mass of the system. This is justified by noting that Kroupa \& Bastian
(1997) found that the space motions of the LMC and SMC are roughly
parallel; 2 - The value for the Draco proper motion given in Scholz \&
Irwin (1994) includes the correction for the solar motion.}
\halign{\hfil#\hfil&\quad\hfil#\hfil&\quad\hfil#\hfil&\quad\hfil#\hfil&\quad\hfil#\hfil&\quad\hfil#\hfil&\quad\hfil#\hfil&\quad\hfil#\hfil&\quad\hfil#\hfil&\quad\hfil#\hfil&\quad\hfil#\hfil&\quad\hfil#\hfil\cr
\noalign{\hrule}
\noalign{\vskip0.1truecm}
Name&$\ell$&$b$&    $s$&    $r$& v$_{\odot}$&
$\mu_{\alpha}\cos\delta$& $\mu_{\delta}$& $v_r$& $v_t$& Type &Source \cr
\tableunderline
$^1$LMC/SMC & 282 & -34 & 49 & 49 & 274 & 0.161 $\pm$ 0.019 & -0.006 $\pm$
0.021 & 83 & 249 & Irr III-IV & 1, 7\cr
\tableunderline
Sculptor& 288  & -83    &     79&            79 &        108&
0.072 $\pm$ 0.022 & -0.006 $\pm$ 0.025& 95 &  202 & dSph & 2, 7\cr
\tableunderline
Ursa Minor& 105 & 45 & 66 & 68 &   -248 & 0.022 $\pm$ 0.008 & 0.026
$\pm$ 0.01 & -87 & 264 & dSph & 3, 7\cr
\tableunderline
NGC4147 & 253 & 77 & 19 & 21 & 183 & -0.27 $\pm$ 0.13 & 0.09 $\pm$
0.13 & 222 & 248 & GC & 4\cr
\tableunderline
Pal 3&     240&  42&     89&            93&        83&     0.033
$\pm$ 0.023 &  0.030 $\pm$ 0.031 & -65 & 353 & GC & 5\cr
\tableunderline
$^2$Draco & 86 & 35 & 82 & 82 & -293 & 0.09 $\pm$ 0.05 & 0.1 $\pm$ 0.05 &
-255 & 454 & dSph & 6, 7\cr
\tableunderline
\noalign{\vskip0.1truecm}\noalign{\hrule}
}
\endtable

\beginfigure{4}
\centerline{\psfig{figure=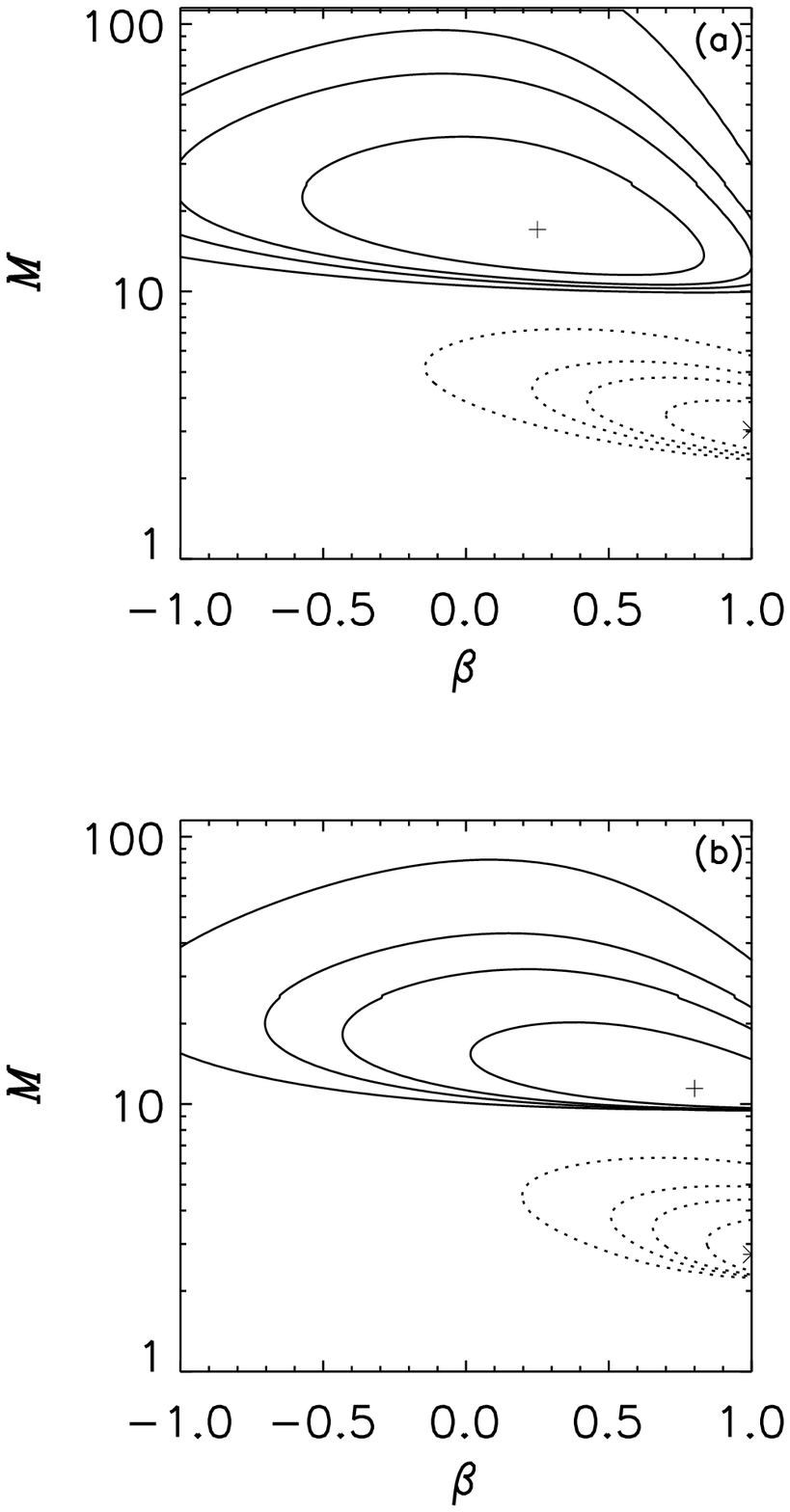,width=\hssize}}
\smallskip\noindent
\caption{{\bf Figure 4} (a) Likelihood contours for the total mass $M$
(in units of $10^{11}M_{\odot}$) and the velocity anisotropy $\beta$
obtained assuming a shadow tracer satellite population with $\as =
100$ and using Milky Way satellite and globular cluster radial
velocities only. Results including Leo~I (solid curves) and excluding
Leo~I (dotted curves) are shown. Contours are at heights of 0.32, 0.1,
0.045 and 0.01 of peak height. (b) As in (a) but for the case of a
power-law tracer satellite population with $\gamma = 3.4$.}
\endfigure

\beginfigure{5}
\centerline{\psfig{figure=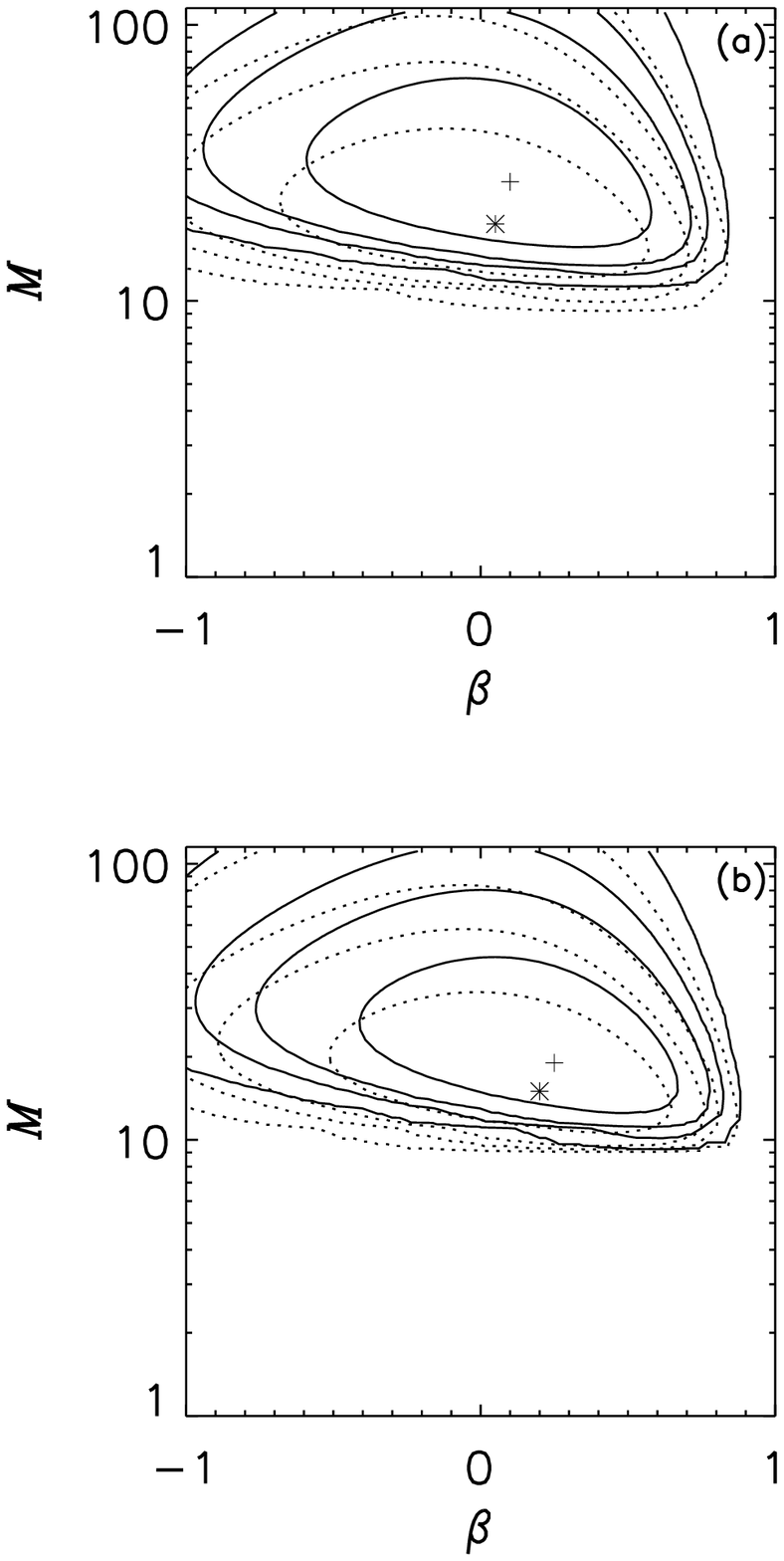,width=\hssize}}
\smallskip\noindent
\caption{{\bf Figure 5} (a) Likelihood contours for the mass $M$
(in units of $10^{11}M_{\odot}$) and the velocity anisotropy $\beta$
obtained assuming a shadow tracer satellite population with $\as =
100$ and using Milky Way satellite and globular cluster radial
velocities and proper motions. Results including Leo~I (solid curves)
and excluding Leo~I (dotted curves) are shown. Contours are at heights
of 0.32, 0.1, 0.045 and 0.01 of peak height. (b) As in (a) but for the
case of a power-law tracer satellite population with $\gamma$ = 3.4.}
\endfigure

In order to find the likelihood of a particular set of model
parameters given the observations of radial velocities and proper
motions, we make use of Bayes' theorem. This gives us the fundamental
formula of the algorithm, namely
\eqnam{\modelprob}
$$P(a,\beta|r_i,v_{ri},I) = {1 \over N} P(a)P(\beta)\Pi_{i=1}^N
P(r_i,v_{ri}|a,\beta),\eqno\new$$
where the normalisation N is given by 
$$N = \int da d\beta P(a) P(\beta|a)\Pi_{i=1}^N
P(r_i,v_{ri}|a,\beta).\eqno\new$$
Here, $P(a,\beta|r_i,v_{ri},I)$ is the probability of the model
parameters taking the values $a$ and $\beta$ given the data
($r_i$,$v_{ri}$). $I$ denotes the prior information, namely the prior
probability distributions, $P(a)$ and $P(\beta)$ respectively. We
initially chose $P(a) \propto 1/a$, as recommended by Kendall \&
Stuart (1977) as a suitable prior for a variable that can take values
within the range ($0,\infty$). We also experimented with $P(a)
\propto 1/a^2$ as this prior gives lower probabilities for very large
(and physically unreasonable) halos. The prior in the velocity
anisotropy is taken to be of the general form 
\eqnam{\priorbeta}
$$P(\beta) \propto 1/(3-2\beta)^{n}.$$
When $n=0$, this is a uniform prior. When $n=2$, this corresponds to
the prior introduced by Kochanek (1986) in which the ratio of radial
kinetic energy to total kinetic energy is uniform. For $n >1$, the
ratio of the probability of obtaining a radial $\beta$ to that of
obtaining a tangential $\beta$ is $3^{n-1}-1$. As $n \rightarrow
\infty$, the prior becomes increasingly biased towards radial
anisotropy. Numerical simulations of halo formation do suggest that
halos may well be radially anisotropic (e.g., Dubinski \& Carlberg
1991).

\section{Results} 
In applying the Bayesian analysis to the observational data, two kinds
of calculations suggest themselves. First, the gas rotation curve may
be a good guide to the velocity normalisation $v_0$ in the distant
halo. Second, the gravity field in the outer parts of the halo may be
very different from the inner parts and so $v_0$ may be unrelated to
the circular velocity near the Sun (c.f. Little \& Tremaine 1987). In
this Section, the velocity normalisation $v_0$ is {\it always} chosen
so that the circular speed at the solar radius is $220$ \kms. Section
4 studies the implications of allowing $v_0$ to vary.

\subsection{The Data on the Satellites}
Tables~2 and~3 summarise the data available on satellites and globular
clusters at distances greater than 20 kpc from the Galactic Centre,
correcting several errors contained in previous presentations of these
data -- tables of the proper motion data in particular have tended to
harbour serious inconsistencies of notation. The radial velocities
quoted for the dwarf spheroidals are all based on optical observations
except for that of Phoenix which is derived from radio measurements.

In converting the heliocentric quantities to Galactocentric ones, we
assume a circular speed of 220 \kms at the Galactocentric radius
of the sun ($R_{\odot}$ = 8.0 kpc) and a solar peculiar velocity of
($U,V,W$) = (-9,12,7), where $U$ is directed outward from the Galactic
Centre, $V$ is positive in the direction of Galactic rotation at the
position of the sun, and $W$ is positive towards the North Galactic
Pole. Heliocentric radial velocities are first corrected for solar
motion using these values and then adjusted by a factor to take
account of contamination of the observed radial velocity by the
(unknown) tangential velocity components. This correction is derived
from the geometric relationship
$$v_{r\odot} = v_r \cos \alpha + v_t\sin \alpha \cos \psi.\eqno\new$$
Here, $v_{r\odot}$ is the observed heliocentric radial velocity, $v_r$
is the Galactocentric radial velocity, $\alpha$ is the angle between
the unit vector $\rhat$ from the Galactic Centre to the satellite and
the unit vector $\shat$ from the sun to the satellite, and $\psi$ is
the angle between the normal to the orbital plane and $\rhat \times
\shat$. As the DFs depend only on the energy and the modulus of the
angular momentum, the velocity ellipsoid is aligned in spherical polar
coordinates. By squaring and averaging over the distribution function,
we find the (statistical) correction factor is
\eqnam{\galactocorr}
$$\langle v_{r}^2 \rangle^{1/2}= {\langle v_{r\odot}^2 \rangle^{1/2} 
\over \sqrt{1 - \beta\sin^2\alpha}},\eqno\new$$
Here, the angled brackets denote ensemble averages, while $\beta$ is
the constant orbital anisotropy. This statistical correction is small
for all the satellites in our dataset, even those at Galactocentric
radii close to 20 kpc where the offset of the line of sight is
greatest.
\begintable*{4}
\def\tableunderline{\noalign{\vskip0.1truecm}\noalign{\vskip0.1truecm}}
\caption{{\bf Table 4.} Mass estimates obtained using Bayesian
analysis applied to the radial velocity data only. All masses are in
units of $10^{11} M_{\odot}$ and all lengths are in kpc.}
\halign{\hfil#\hfil&\quad\hfil#\hfil&\quad\hfil#\hfil&\quad\hfil#\hfil&\quad\hfil#\hfil&\quad\hfil#\hfil&\quad\hfil#\hfil&\quad\hfil#\hfil&\quad\hfil#\hfil\cr
\noalign{\hrule}
\noalign{\vskip0.1truecm}
&&&&Shadow&Tracers&&&\cr
$\as$ & $a$ prior & $\beta$ prior & & Most likely $\beta$& Most likely
$a$& Most likely $M_{tot}$& $M(<50)$& $M(<100)$\cr
\tableunderline
100&1/$a^2$&Energy&With Leo~I&0.25&150&17.0&5.3&9.4\cr
&	&	&Without Leo~I&1.0&25&3.0&2.6&2.9\cr
\tableunderline
100&1/$a$&Energy&With Leo~I&0.15&180&20.5&5.4&9.8\cr
&	&	&Without Leo~I&1.0&25&3.0&2.6&2.9\cr
\tableunderline
100&1/$a^2$&Uniform&With Leo~I&-0.1&175&19.5&5.4&9.8\cr
&	&	&Without Leo~I&0.95&25&3.0&2.6&2.9\cr
\tableunderline
$a_{\rm{halo}}$&1/$a^2$&Energy&With Leo~I&0.2&135&15.0&5.3&9.1\cr
&	&	&Without Leo~I&1.0&36&4.1&3.4&3.9\cr
\tableunderline
\noalign{\vskip0.1truecm}\noalign{\hrule}
\noalign{\vskip0.1truecm}
&&&&Power-Law&Tracers&&&\cr
\noalign{\vskip0.1truecm}
$\gamma$ & $a$ prior & $\beta$ prior & & Most likely $\beta$& Most likely
$a$& Most likely $M_{tot}$& $M(<50)$& $M(<100)$\cr
\noalign{\vskip0.1truecm}
3.4&1/$a^2$&Energy& With Leo~I&0.8&100&11.4&5.0&8.0\cr
&	&	&Without Leo~I&1.0&23&2.7&2.5&2.7\cr
\tableunderline
3.4&1/$a$&Energy& With Leo~I&0.75&105&12.0&5.1&8.2\cr
&	&	&Without Leo~I&1.0&24&2.8&2.6&2.8\cr
\tableunderline
3.4&1/$a^2$&Uniform& With Leo~I&0.35&120&13.5&5.2&8.7\cr
&	&	&Without Leo~I&1.0&23&2.7&2.5&2.7\cr
\tableunderline
4&1/$a^2$&Energy& With Leo~I&1.0&105&12.0&5.1&8.2\cr
&	&	&Without Leo~I&1.0&28&3.3&2.9&3.2\cr
\tableunderline
\noalign{\vskip0.1truecm}\noalign{\hrule}
}
\endtable
\subsection{Results with Radial Velocity Data Only}
Let us now apply the methods described in Section~2 to the
observational data. In order to emphasise the crucial role played by
the proper motions, we first present the results obtained using only
the radial velocities of the satellites. Fig.~4(a) shows the
likelihood contours in the mass--anisotropy ($M$-$\beta$) plane for
the case of a shadow tracer satellite population with $\as$ = 100
kpc. The contours obtained when Leo~I is assumed to be bound to the
Milky Way (solid contours) are very different from those obtained when
Leo~I is excluded from the dataset (dotted contours). The maximum of
the probability surface is shown as a cross (asterisk) for the
contours including (excluding) Leo~I. When Leo~I is included, the most
likely value of the total halo mass $M$ is
17.0$\times10^{11}$M$_{\odot}$ corresponding to a scalelength $a$ of
150 kpc for the halo. When Leo~I is excluded, the most likely values
of $M$ and $a$ shrink to 3.0$\times10^{11}$M$_{\odot}$ and 25 kpc
respectively. The contours in Fig.~4(a) are obtained using $1/a^2$ as
the prior probability on $a$ and the uniform energy prior ($n=2$) on
$\beta$. Fig.~4(b) shows the contours obtained using the same priors
but for a power-law tracer satellite population with $\gamma$ =
3.4. In this case, the most likely value of $M$ is
11.4$\times10^{11}$M$_{\odot}$ ($a$ = 100 kpc) if Leo~I is included
and 2.7$\times10^{11}$M$_{\odot}$ ($a$ = 25 kpc) if Leo~I is
excluded. For both shadow and power-law tracer populations, we
conclude that if we use only radial velocities to estimate the total
mass of the Milky Way halo, then the dominant uncertainty is whether
or not Leo~I is bound.

Table~4 summarises the results obtained for both shadow tracers and
power-law tracers using a variety of different priors on $a$ and
$\beta$. This table also illustrates the effect of varying the assumed
value of $\as$ for a shadow tracer population and $\gamma$ for a power
law tracer population.  There are a number of trends visible in the
results of Table 4. We observe that for the shadow tracers, changing
the prior on $a$ from $1/a$ to $1/a^2$ leads to a decrease in the
estimate of the total mass.  This is natural, since by choosing the
$1/a^2$ prior we are forcing the halo to be smaller. Exactly the same
effect is observed for the power-law tracers.  Changing the prior on
$a$, however, has the desirable effect of reducing the size of the
likelihood contours in the ($M$-$\beta$) plane. This has a sound
physical basis, as the Milky Way halo cannot extend to Megaparsec
scales (see e.g., Evans (1997), Gates, Kamionkowski \& Turner (1997),
as well as Cowsik, Ratnam \& Bhattacharjee (1996) for a heterodox
viewpoint). Switching the prior on $\beta$ from the uniform energy
prior ($n=2$) to a uniform prior ($n=0$) leads to an increase in the
mass estimates including Leo~I. This may be understood by noting that
the uniform energy prior is biased towards radially anisotropic
velocity distributions. A uniform prior gives comparatively more
weight to tangential distributions in which the satellites have large
(unknown) tangential velocities. This, naturally, implies a larger
total halo mass.

Table 4 also shows that our choice of $\as$ for the shadow tracers
does not have a significant effect on the mass estimate. If, instead
of using $\as$ = 100 kpc, we assume that the satellite scalelength is
the same as that of the halo, the mass estimate both with and without
Leo~I are changed by less than $30$ \%. For the power-law tracers,
increasing the value of the power index $\gamma$ leads to an increase
in the mass estimate. This may be understood in terms of the
likelihood of obtaining a distant satellite in a power-law density
model. As $\gamma$ increases, the satellite density falls off faster,
making distant satellites less common. In order to fit the observed
data which contains distant satellites, the halo must necessarily be
larger.

\begintable*{5}
\def\tableunderline{\noalign{\vskip0.1truecm}\noalign{\vskip0.1truecm}}
\caption{{\bf Table 5.} Mass estimates obtained using Bayesian
analysis applied to the radial velocity and proper motion data,
assuming a Lorentzian error convolution function for the observational
errors on the proper motions. All masses are in units of $10^{11}
M_{\odot}$ and all lengths are in kpc.}
\halign{\hfil#\hfil&\quad\hfil#\hfil&\quad\hfil#\hfil&\quad\hfil#\hfil&\quad\hfil#\hfil&\quad\hfil#\hfil&\quad\hfil#\hfil&\quad\hfil#\hfil&\quad\hfil#\hfil\cr
\noalign{\hrule}
\noalign{\vskip0.1truecm}
&&&&Shadow&Tracers&&&\cr $\as$ & $a$ prior & $\beta$ prior & & Most
likely $\beta$& Most likely $a$& Most likely $M_{tot}$& $M(<50)$&
$M(<100)$\cr
\tableunderline
100&1/$a^2$&Energy&With Leo~I&0.1&240&27.0&5.5&10.4\cr
&	&	&Without Leo~I&0.05&170&19.0&5.4&9.7\cr
\tableunderline
100&1/$a$&Energy&With Leo~I&0.05&295&33.0&5.5&10.7\cr
&	&	&Without Leo~I&0.0&205&23.0&5.5&10.1\cr
\tableunderline
100&1/$a^2$&Uniform&With Leo~I&-0.15&260&29.0&5.5&10.4\cr
&	&	&Without Leo~I&-0.2&185&21.0&5.4&9.9\cr
\tableunderline
$a_{\rm{halo}}$&1/$a^2$&Energy&With Leo~I&-0.15&240&27.0&5.5&10.4\cr
&	&	&Without Leo~I&-0.1&170&19.0&5.4&9.6\cr
\tableunderline
\noalign{\vskip0.1truecm}\noalign{\hrule}
\noalign{\vskip0.1truecm}
&&&&Power-Law&Tracers&&&\cr
\noalign{\vskip0.1truecm}
$\gamma$ & $a$ prior & $\beta$ prior & & Most likely $\beta$& Most likely
$a$& Most likely $M_{tot}$& $M(<50)$& $M(<100)$\cr
\noalign{\vskip0.1truecm}
3.4&1/$a^2$&Energy& With Leo~I&0.25&170&19.0&5.4&9.6\cr
&	&	&Without Leo~I&0.2&135&15.0&5.3&9.1\cr
\tableunderline
3.4&1/$a$&Energy& With Leo~I&0.15&225&25.0&5.5&10.3\cr
&	&	&Without Leo~I&0.1&170&19.0&5.4&9.6\cr
\tableunderline
3.4&1/$a^2$&Uniform& With Leo~I&0.0&205&23.0&5.5&10.1\cr
&	&	&Without Leo~I&-0.05&150&17.0&5.3&9.1\cr
\tableunderline
4&1/$a^2$&Energy& With Leo~I&0.3&205&23.0&5.5&10.1\cr
&	&	&Without Leo~I&0.3&150&17.0&5.3&9.1\cr
\tableunderline
\noalign{\vskip0.1truecm}\noalign{\hrule}
}
\endtable

\beginfigure{6}

\centerline{\psfig{figure=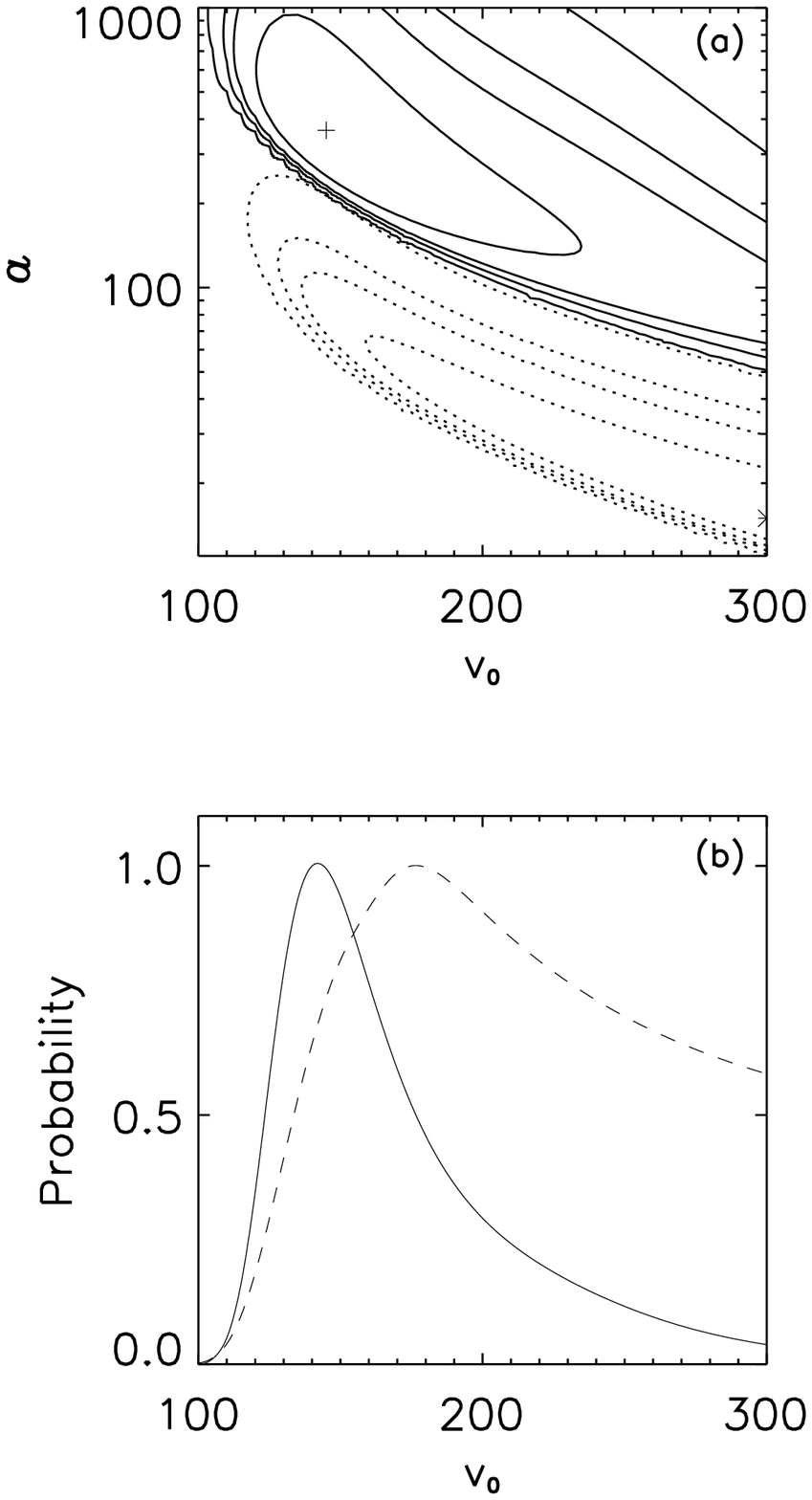,height=2\hssize}}

\smallskip\noindent
\caption{{\bf Figure 6.} Likelihood contours for the scale-length $a$
(in kpc) and the velocity normalisation $v_0$ obtained assuming a
shadow tracer satellite population with $\as = 100$ and using Milky
Way satellite and globular cluster radial velocities only. Results
including Leo~I (solid curves) and excluding Leo~I (dotted curves) are
shown. Contours are at heights of 0.32, 0.1, 0.045 and 0.01 of peak
height. Also shown is the marginal distribution for the velocity
normalisation. The contours are generated assuming the uniform energy
prior probability for $\beta$ and a $1/v_0^2$ prior on $v_0$.}
\endfigure

\beginfigure{7}

\centerline{\psfig{figure=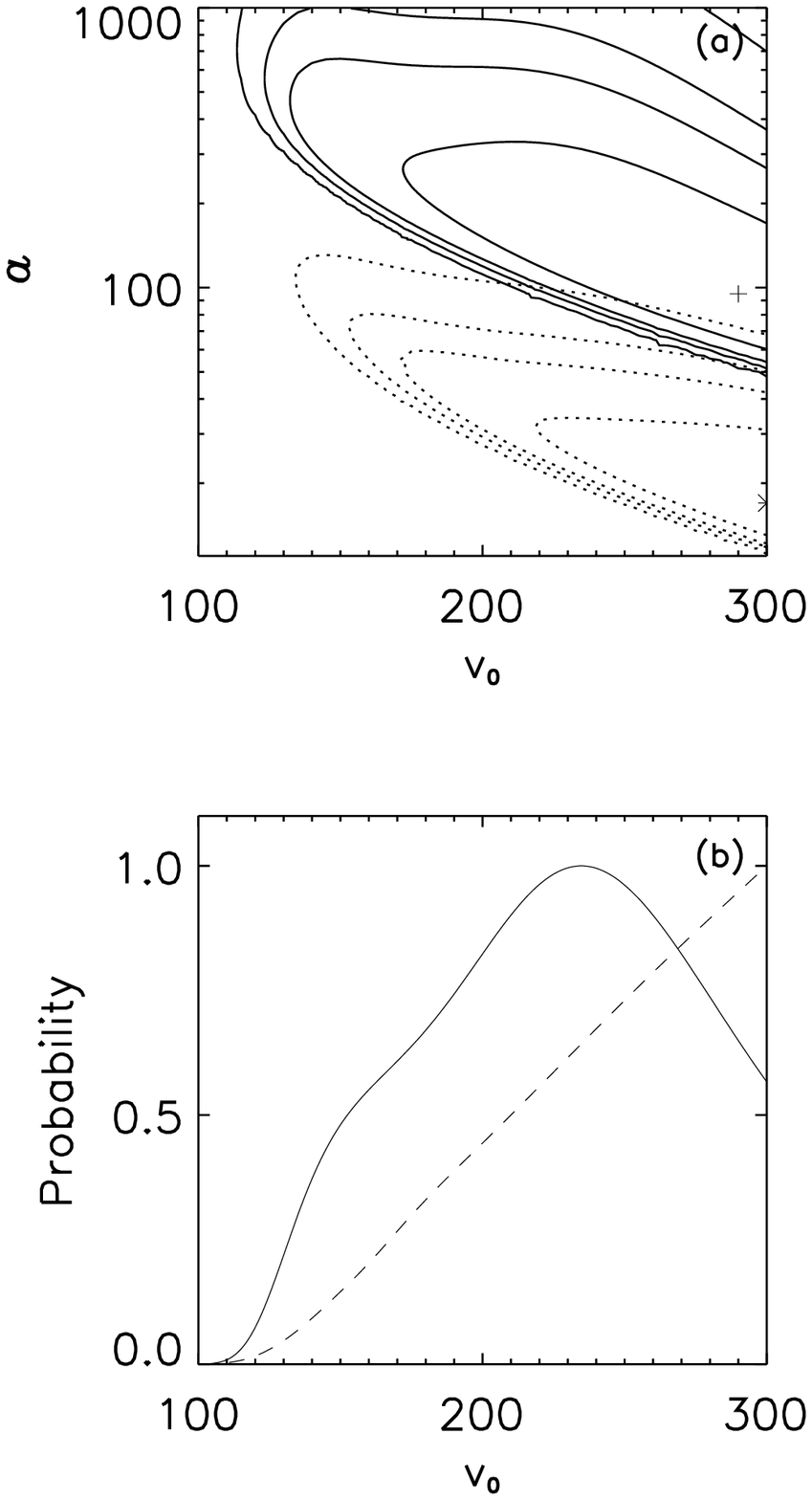,height=2\hssize}}

\smallskip\noindent
\caption{{\bf Figure 7.} As Fig.~6, but the contours are generated 
assuming uniform prior probabilities for both $\beta$ and $v_0$.}
\endfigure

\beginfigure*{8}

\centerline{\psfig{figure=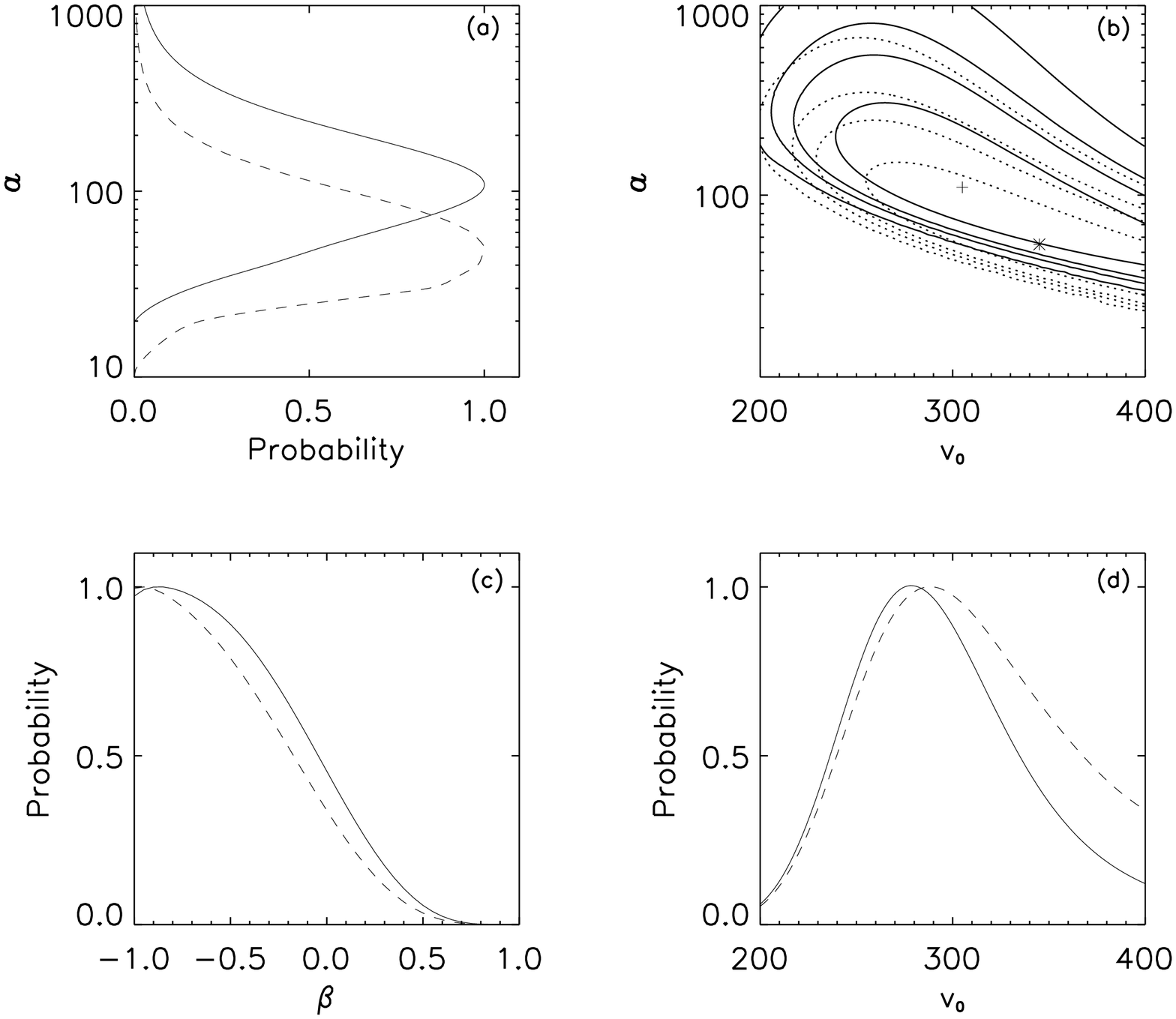,height=2\hssize}}

\smallskip\noindent
\caption{{\bf Figure 8.} Likelihood contours for the scale-length $a$
(in kpc) and the velocity normalisation $v_0$ (in \kms) obtained
assuming a shadow tracer satellite population with $\as = 100$ and
using Milky Way satellite and globular cluster radial velocities and
proper motions. Results including Leo~I (solid curves) and excluding
Leo~I (dotted curves) are shown. Contours are at heights of 0.32, 0.1,
0.045 and 0.01 of peak height. Also shown are the marginal
distributions for the three parameters, including the velocity
anisotropy $\beta$.}
\endfigure

\beginfigure{9}

\centerline{\psfig{figure=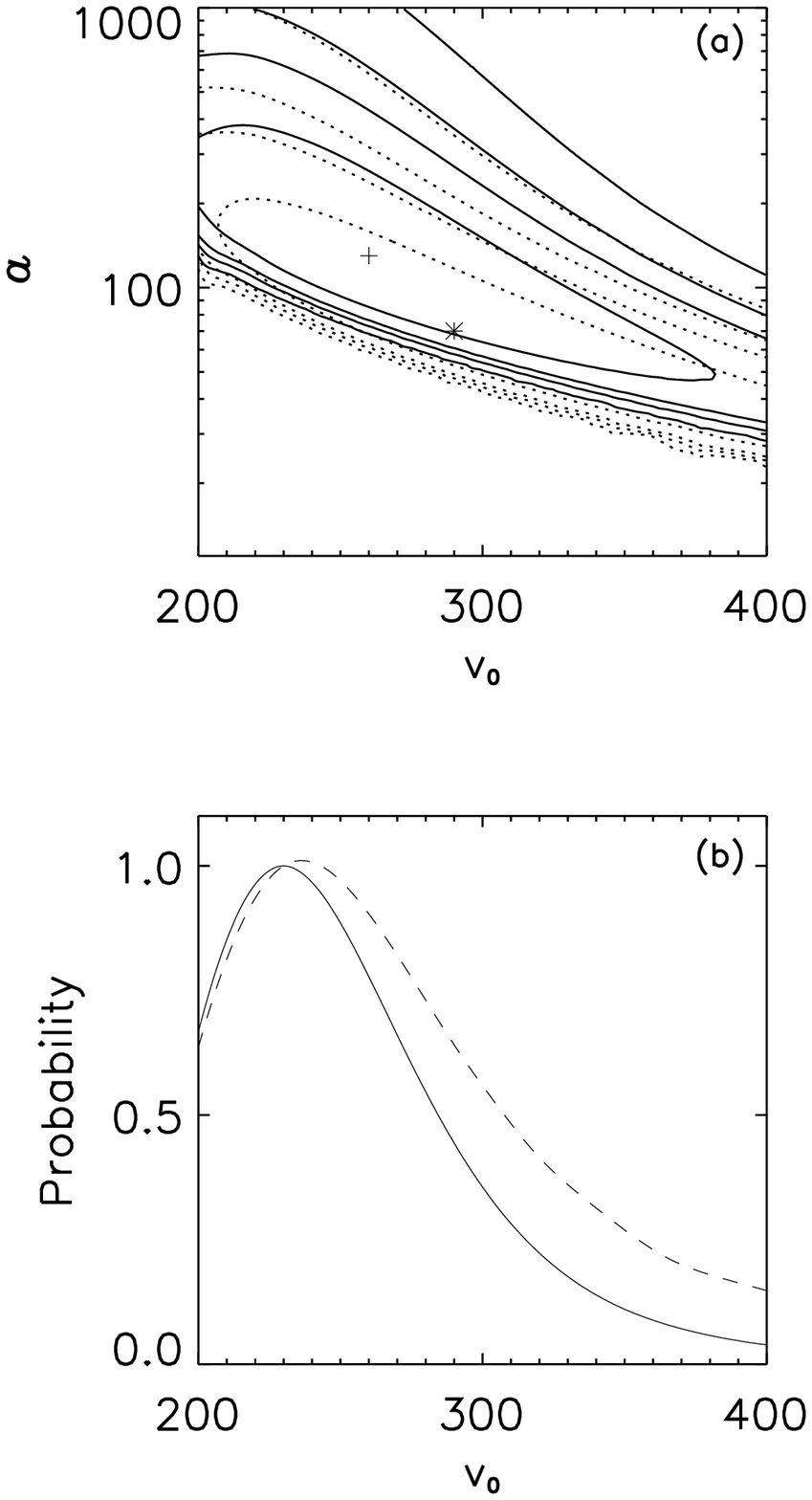,height=2\hssize}}

\smallskip\noindent
\caption{{\bf Figure 9.} Likelihood contours for the scale-length $a$
(in kpc) and the velocity normalisation $v_0$ obtained assuming a
shadow tracer satellite population with $\as = 100$ and using Milky
Way satellite and globular cluster radial velocities and proper
motions. Results including Leo~I (solid curves) and excluding Leo~I
(dotted curves) are shown. Contours are at heights of 0.32, 0.1, 0.045
and 0.01 of peak height. Also shown is the marginal distribution for
the velocity normalisation. The contours were generated assuming a
prior probability for $\beta$ which is strongly biased towards radial
anisotropy (see text for discussion).}
\endfigure

\beginfigure{10}

\centerline{\psfig{figure=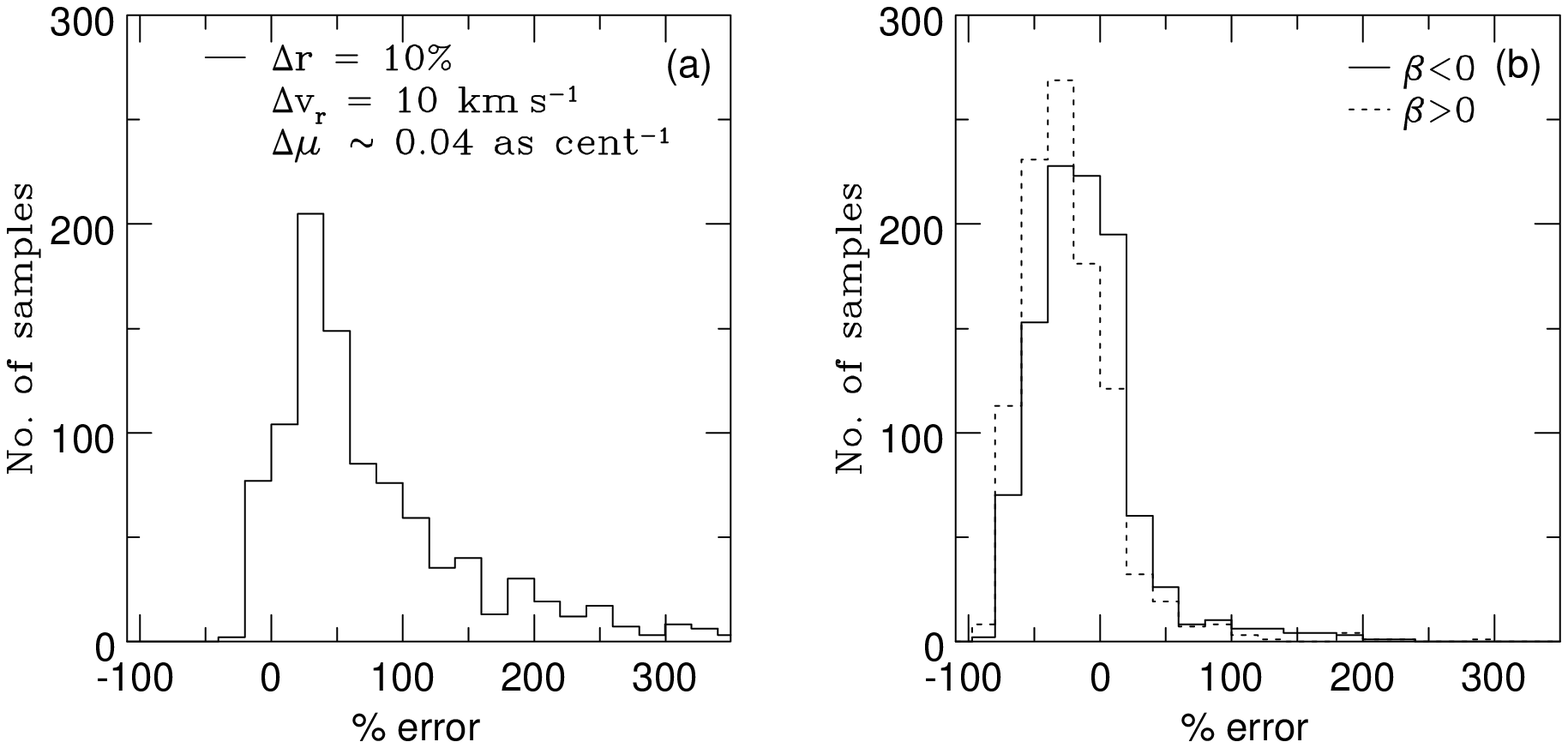,height=.5\hssize}}

\smallskip\noindent
\caption{{\bf Figure 10.} (a) Histogram showing the effects of
present-day measurement errors on the mass estimate obtained. (b)
Histograms illustrating the effects of streams in the data, when all
30 data-points lie on either of two streams. In both cases, the
histograms show the number out of 1000 datasets yielding a given
percentage error in the mass estimate $M$.}
\endfigure

\beginfigure{11}
\centerline{\psfig{figure=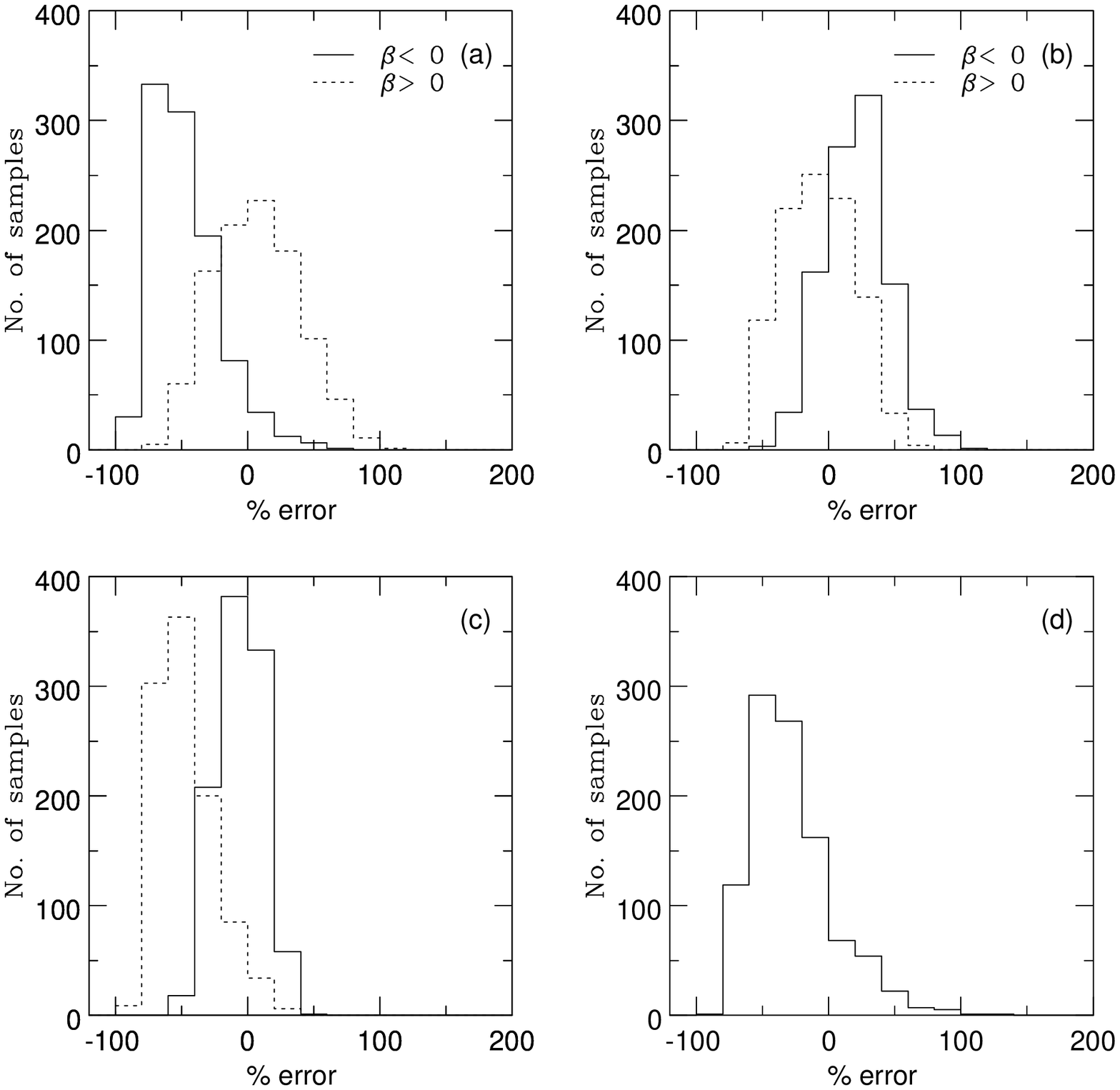,height=\hssize}}
\smallskip\noindent
\caption{{\bf Figure 11.} Histograms to illustrate the effects of
various modelling uncertainties on the mass estimate. All histograms
show the number of datasets out of 1000 which yielded a given
percentage error in $M$. (a) Datasets of 30 points using only radial
velocities -- lack of knowledge about $\beta$ gives uncertainty in
mass (b) As in (a) but with proper motions included. (c) Datasets of
30 points generated using a TF model with $\as$=100 kpc but with a
power-law model (with $\gamma = 4.0$) used in the Bayesian
analysis. Two cases are shown, with (solid curve) and without (dashed
curve) proper motion data. (d) Datasets of 30 radial velocities where
the velocity normalisation is allowed to be a free parameter in the
algorithm. The uncertainty in $M$ is not significantly increased above
that in Fig. 12(a) in which $v_0$ is assumed known.}
\endfigure

\subsection{Results with Radial and Proper Motion Data}

Having considered the radial velocity data in isolation, we now
include the available proper motions in our analysis. In the past few
years, the number of measured proper motions of distant clusters and
satellites has doubled and the accuracy of these measurements has
improved. More importantly, the future holds the prospect of rapid
progress using space-based astrometric satellites.  At present, the
proper motion errors are still large and so we must take account of
them. This is done by convolving the probabilities given in Section~2
with an error function to obtain the probability
$P(r_i,v_{i,obs}|a,\beta)$ of obtaining the observed full-space
velocity $v_{i,obs}$ given the values of the model parameters. Thus we
obtain
\eqnam{\convolprob}
$$\eqalign{P(r_i,v_{i,obs}|a,\beta) = \int\int dv_{\alpha}&dv_{\delta}
E_1(v_{\alpha})E_1(v_{\delta})\cr &\times P(r_i,v_{i}(v_{\odot},v_{\alpha},v_{\delta})|a,\beta),}\eqno\new$$
where $v_{\alpha}$ and $v_{\delta}$ are the velocities perpendicular
to the line of sight and $v_\odot$ is the radial velocity. The error
convolution function $E_1(v_{\alpha})$ is the probability of obtaining
the observations given the true velocity $v_{\alpha}$ and the
estimates of the associated errors.  It is likely that the
observational errors are strongly non-Gaussian. We assume the
Lorentzian error convolution function $E_1$ given by
$$E_1(v) = {1 \over
\sqrt{2}\pi\sigma_1}{2\sigma_1^2 \over 2\sigma_1^2 +
(v-v_{obs})^2}.\eqno\new$$
Lorentzians have broader wings than the more familiar Gaussians. In
fact, $E_1(v)$ is the first member of a sequence of error convolution
functions which gradually tend towards Gaussianity. As this family of
functions may find further applications in astronomy, their properties
are presented in more detail in Appendix A. Here, we note only that in
order to normalise $E_1$, we choose $\sigma_1$ such that the quartiles
of $E_1$ are the same as those of a Gaussian of width $\sigma_{G}$,
where $\sigma_{G}$ is the published error estimate. We obtain the
relation
$$ \sigma_1 = 0.477\sigma_{G}\eqno\new$$
and use this in all the convolutions. In what follows we neglect
the errors in the heliocentric radial velocities and distances of the
satellites, as initial tests indicated that their effect is negligible
compared to that of the proper motions.

Fig.~5 shows the likelihood contours obtained by the above procedure
for our standard shadow tracer and power-law tracer models. {\it There
is good agreement between the contours based on datasets with and
without Leo~I}. This removes a longstanding impasse in the subject
(c.f., Little \& Tremaine 1987; Kulessa \& Lynden-Bell 1992; Kochanek
1996).

Table~5 summarises the results obtained when the proper motions of the
satellites are included. It can be compared directly with Table~4
obtained for the same models and priors. As before, changing the $a$
prior from $1/a$ to $1/a^2$ leads to a decrease in the mass
estimate. Changing the assumed value of $\as$ for the shadow tracer
population leaves the mass estimate unchanged, although $\beta$ moves
towards more tangential velocity distributions. Increasing the value
of $\gamma$ for the power-law tracers increases the mass estimates
both with and without Leo~I by $\lta$ 20\%.

To determine our best estimates for the mass, we compare the area of
overlap of the contours with and without Leo~I for each of the models
in Table~5 and calculate the fractional change in the mass estimate
when Leo~I is removed. This area is maximised and the fractional
change minimised when the total mass of the halo is
2.7$\times10^{12}$M$_{\odot}$ assuming a shadow tracer population with
scalelength 100 kpc, and 1.9$\times10^{12}$M$_{\odot}$ assuming a
power-law tracer population with $\gamma$ = 3.4.  These two mass
estimates are in reasonably good agreement, in that the estimate
obtained from the power-law tracers lies comfortably within the $32\%$
contour for the shadow tracers and vice versa. The power-law tracer
result is more insensitive to the presence of Leo~I and we therefore
conclude that it is (marginally) the more satisfactory of the two. The
$32\%$ contours in Fig.~5(b) give a range of $1.5$ - $4.8$
$\times10^{12}$M$_{\odot}$ for the power-law estimate. As we shall see
in Section~5, this range is in fact an underestimate of the
true uncertainty.

It is interesting at this point to ask how likely it is that a single
satellite in a dataset of 30 objects with radial velocities drawn
randomly from our TF halo model would make a substantial difference to
the mass estimate. We generate 1000 datasets and obtain the likelihood
contours with the full dataset and minus each of the satellites in
turn. We find that approximately $0.3\%$ of datasets contain a
satellite for which the mass estimates with and without the satellite
differ by more than a factor of five (c.f. Fich \& Tremaine
1991). This result varies from $0.1-0.5\%$ as $\beta$ is varied from
$0.9$ to $-9.0$. Thus, Leo~I is a rather unusual object and our prior
expectation is not to find such a satellite.  In the simulations, the
data are generated consistently from the model, but it is nonetheless
the case that removing one data point from the radial velocity dataset
can very occasionally cause a large shift in the likelihood contours
in the Bayesian analysis.

\section{The Velocity Normalisation}
Thus far, our analysis has assumed that the normalisation $v_0$ of our
halo model is fixed by the constraint that the rotation curve has an
amplitude of $\sim 220$ \kms at the Sun. We regard this as an
economical assumption to make. For example, if an isotropic tracer
population has a density falling off like $\rho \sim r^{-3}$ in a
galaxy with a flat rotation curve of amplitude $v_0$, then
(Lynden-Bell \& Frenk 1981, Evans, H\"afner \& de Zeeuw 1997)
$$v_0^2 = 3 \langle v_r^2 \rangle.$$
Using the data in Table 2 and 3, this gives $v_0 = 220$
\kms almost exactly -- in good agreement with our assumption.
Nonetheless, it is clearly interesting to relax this condition,
especially as visible matter dominates the gravity field at the solar
radius whilst the dwarf satellites are in the region where dark matter
dominates. As Little \& Tremaine (1987) first pointed out, the fact
that rotation curves at radii between 5 and 20 kpc indicate a
logarithmic potential, and that satellite galaxies at 100 kpc indicate
a logarithmic potential, does not imply that the circular speeds in
the two potentials are precisely the same.  So, this section presents
results when the Bayesian analysis is performed in the three
dimensional parameter space of $a$, $\beta$ and $v_0$. The required
probabilities are easily obtained by a straightforward extension of
the formulae in Section 2.

Fig.~6 (a) presents contours in the ($a,v_0$) plane for a shadow
tracer model when only the radial velocities of the satellites are
used. In generating this figure, a prior probability of $1/v_0^2$ was
assumed for $v_0$ and the uniform energy ($n=2$) prior was used for
$\beta$. As in Fig.~4 (a) the contours with and without Leo~I are
disjoint at the 99\% level. The marginal distributions of Fig.~6 (b)
show that the most likely values of $v_0$ are $140$ \kms including
Leo~I and $175$ \kms excluding Leo~I. We also find that $\beta \sim 1$
both with and without Leo~I.  This seems in accord with the earlier
results of Little \& Tremaine (1987), who estimated $v_0 \lta 165$
\kms using a smaller sample of ten objects together with an infinite
isothermal sphere model.  The situation changes dramatically, however,
if we change the prior probabilities used in the Bayesian
analysis. Fig.~7 presents the likelihood contours and marginal
distributions for $v_0$ when the priors on $v_0$ and $\beta$ are now
uniform. As the figure clearly shows, the most likely values for $v_0$
are now $235$ \kms with Leo~I and $>300$ \kms without. This strong
sensitivity to the choice of priors is a cause for alarm, and suggests
that none of the values of the velocity normalisation are firmly
established. Let us also remark that the likelihood curves in the
plane of ($M,\beta$) with and without Leo~I always remain disjoint. In
fact, if we simply adjust the value of the velocity normalisation with
the aim of obtaining overlapping contours in the ($M,\beta$) plane, we
are driven towards extremely low values of $v_0$ ($\sim 80$ \kms) and
unphysically large values of the scalelength $a$ ($\sim 1$ Mpc). This
is simply because reducing the value of $v_0$ means that a larger halo
scalelength is required to ensure that all the satellites are bound --
for large scalelengths, Leo~I is buried deep within the halo and
therefore has a less significant effect on the total mass.

We now proceed to include the proper motion data to see if the
situation is improved.  Fig.~8 (b) presents contours in the ($a,v_0$)
plane for a shadow tracer model when the proper motion data are
included, together with the marginal distributions for the three model
parameters. We note first that the most likely values of $a$ and $v_0$
yield a mass estimate of $2.4 \times 10^{12}$ M$_{\odot}$ with Leo~I
and $1.7 \times 10^{12}$ M$_\odot$ without Leo~I. This is in broad
agreement with the results of Section 3.  However, it is clear that
the details of the results in Fig.~8 are significantly different. In
particular, the anisotropy parameter now has a most likely value of
$-0.95$, indicating a strongly tangential velocity distribution.
Perhaps more significantly, the marginal distributions in Fig.~8
indicate that most likely value for the velocity normalisation is
$\sim 280$ \kms, irrespective of whether Leo~I is included or
excluded.  Our high result for the velocity normalisation is an
inevitable consequence of the tangential anisotropy of the subsample
of 6 satellites with proper motions, which already have $\beta \sim
-1$. The greater the tangential anisotropy, the larger the
normalisation of the halo required to confine the satellites.  There
are worries about the credibility of this velocity normalisation, as
the subsample of satellite galaxies with proper motions may suffer
from selection effects.  First, it is evidently easier to measure the
proper motions for the closer satellites. If the velocity anisotropy
changes in a systematic way -- for example, if it becomes more
radially anisotropic with radius -- then the closer satellites will
not be representative.  Second, the large errors in some of the
present proper motions mean that even the direction of the proper
motions is sometimes in doubt as the error exceeds the absolute value
of the measurement. For example, in the case of Sculptor, the best
value of $\mu_\delta$ is nearly zero, but the large error can produce
velocities in either direction. Obviously, these will bias our result
towards tangential anisotropy and higher velocity normalisation.  A
third selection effect is the way in which the objects for which we
currently have proper motion data were chosen. As described in
Majewski \& Cudworth (1993), astrometric projects to measure proper
motions are currently `` at the mercy of the interests of earlier
observers'' as this determines whether or not sufficient past epoch
plate material exists for comparison with present epoch plates. This
represents a bias which is almost impossible to model.

More worryingly, Fig.~9 shows the effects of changing the prior on the
velocity anisotropy. Here, we have chosen the $n=10$ case in
(\priorbeta) which means that there is a rather significant bias
towards radial anisotropy.  The upper panel again shows the likelihood
contours in the ($a, v_0$) plane, whilst the lower panel shows the
marginal distribution for $v_0$. The maxima of the marginal
distributions occurs at $v_0 =230$ \kms and at $\beta = 0.5$. Thus by
varying the assumed prior on $\beta$, our best estimates for $v_0$ can
change very considerably, even when the proper motion data are
included. Strong sensitivity to the choice of prior probabilities is a
sign that we are trying to extract too much information from the
available data and we therefore conclude that given the current
dataset it is not realistic to constrain the velocity normalisation of
our model from the satellite data alone.  More optimistically, we will
show in Section~6 that the forthcoming space-borne astrometry
satellites SIM and GAIA will be able to put tight constraints on the
value of $v_0$.

\section{Error Analysis}
This section considers three sources of error -- namely, measurement
errors, uncertainties caused by correlations and streaming in the
satellite galaxies and uncertainties due to the modelling itself. The
fourth major source of uncertainty is that due to the small size of
the dataset -- this will be considered in Section~6.  We use Monte
Carlo simulations to estimate the importance of each effect. We
generate artificial datasets of positions and velocities drawn from
DFs with $a$, $v_0$ and $\beta$ fixed. The algorithm of Section~2 is
then applied to see which values of the model parameters are
recovered, and hence the likely uncertainty caused by the error
source. In most cases, the value of $v_0$ is fixed in the Bayesian
analysis so that the circular speed at the radius of the Sun is $220$
\kms. However, when we consider modeling uncertainties in Section~5.3,
we present the effect of allowing $v_0$ to vary in the analysis. For
each type of uncertainty, this procedure is carried out for 1000
datasets. Let us note that in generating the artificial satellites, we
approximate the anisotropic DFs by Gaussians whose widths are given by
the velocity dispersions. Such approximate DFs can very occasionally
generate objects which are not bound to the Milky Way. We test for
this in each dataset and discard any object that is unbound. Our
approximate DFs still slightly overestimate the number of weakly bound
objects, and this leads to a slight but systematic overestimate of the
mass.  This effect can be observed by careful scrutiny of some of the
histograms presented below (for example, in Fig.~12 (a)). Let us
emphasise that this systematic error is always much smaller than the
errors caused by the effects we are investigating in this section. In
what follows, we always quote uncertainties in terms of the average
absolute deviation about the mean rather than the standard
deviation. This is the preferred way of reporting errors in cases
where the distribution is broad (see e.g., Press, Teukolsky,
Vetterling \& Flannery 1992).

\subsection{Measurement Errors}

\noindent
The typical errors in the radial velocity measurements are $\pm 10$
\kms while those in the heliocentric distances are $\sim 10 \% $. To
determine the importance of these errors for a dataset containing $30$
points, we generate an artificial dataset containing 30 data points
with both radial velocities and proper motions. To simulate crudely
the selection effects in measuring proper motions, we keep only the
largest proper motions, which leaves us with 9 proper motions for the
case illustrated in Fig.~10 (a). We analyse this dataset to obtain a
mass estimate. Fig.~10(a) shows the spread in mass estimates obtained
from $1000$ further realisations of the same dataset generated by
drawing points from Gaussian distributions centred on the data points
and with widths representing the measurement uncertainties. The
assumed errors are $10 \%$ in the distances and $\pm 10$ \kms in the
radial velocity and $\pm 40 \muay$ in the proper motions. For these
values, $\sim 30\%$ of mass estimates lie more than a factor of two
above the value that would be reported in the absence of measurement
errors. The average absolute deviation of the estimates about the mean
value is $90 \%$ of $M$, indicating a very large spread. We conclude
that {\it at present measurement errors are a serious source of
uncertainty in our mass estimate, with the proper motion errors
dominating, giving rise to slightly more than a factor of 2
uncertainty}.

\subsection{Correlations in the Dataset}

\noindent
The use of a Bayesian statistical argument implicitly assumes that the
data constitute a random sampling of the underlying distribution.  As
has been known for some time, a number of the satellites of the Milky
Way appear to lie on one of two great circles (see for example
Lynden-Bell 1976, Kunkel \& Demers 1976, Fusi Pecci, Bellazzini,
Cacciari \& Ferraro 1995). If the satellites are in fact the remains
of larger galaxies which have been torn apart by the tidal forces of
the Milky Way, then their motions will necessarily be highly
correlated. This will reduce the amount of information contained in
the dataset of satellite positions and velocities (although see
Johnston, Zhao, Spergel \& Hernquist (1999) for a possible application
of streams for mass estimates).

To determine whether streams or moving groups of satellites have a
significant effect on our analysis, we generate datasets in the
(somewhat extreme) case in which the satellites are dispersed onto two
streams. Each dataset contains 30 data points and we assume that the
full space velocities of all 30 points are known. The results are
presented in Fig.~10 (b) for halo models with both radially and
tangentially anisotropic velocity distributions. If the satellite
galaxies do lie on streams, then this causes a systematic
underestimate in both the halo length scale $a$ and the mass $M$. The
effect is less significant -- but still present -- when the velocity
dispersion tensor is tangential ($\beta <0$) as opposed to radial
($\beta >0$). The underestimate in the mass is of order $20-50\%$. The
average deviation in the mass estimates about the mean value is $26\%$
of $M$ for $\beta >0$ and $29\%$ for $\beta < 0$. As a result, we
conclude that this source of error is almost certainly not so serious
for our dataset as that caused by measurement errors.

\subsection{Modelling Uncertainties}

\noindent
A third major problem arises from our use of parametric fitting --
there is of course no guarantee that any of our tracer population
densities in our assumed halo models are exact representations of the
satellites in the outer parts of the Milky Way, though Fig.~2 assures
us that they are surely not grossly wrong. Modelling uncertainty could
be minimised by the use of non-parametric fitting as advocated by
Merritt and co-workers, although this would only be advantageous with
a significantly larger dataset (see e.g., Merritt \& Tremblay 1993).

The first major modelling uncertainty stems from our ignorance of the
velocity anisotropy of the satellites, or equivalently the value of
$\beta$. To investigate this, datasets are generated for known
anisotropies and the Bayesian analysis is then applied assuming no
knowledge of $\beta$. The histograms in Figs.~9 (a) and (b) show the
spread in mass estimates obtained from samples of 30 data points with
$\beta > 0$ and $\beta < 0$ both without and with proper motion
data. Using only radial velocities, tangentially anisotropic ($\beta <
0$) velocity distributions cause underestimates in the mass as the
unknown tangential velocities of the satellites are, on average,
greater than their known radial velocities. This effect is really a
consequence of our assumed prior, which favours radial anisotropy.  It
is absent for radially anisotropic velocity distributions, as the
broken curve in Fig.~11 (a) demonstrates. Fig.~11 (b) shows that the
inclusion of proper motions removes this problem. Now, there is a mild
tendency to overestimate the mass by at most $\sim 30\%$ for the
tangentially anisotropic case.  On comparing the separation of the
peaks in the distributions for $\beta >0$ and $\beta <0$ in Figs.~9
(a) and (b), we find that it is $\sim 80\%$ of $M$ when only radial
velocities are used, and reduces to $\sim 40\%$ of $M$ when proper
motion data are included. This is a typical measure of the error
caused by uncertainty in the velocity distributions. We conclude that
this is a serious source of uncertainty, though not as problematic as
the measurement errors.

Fig.~11 (c) shows how the use of an incorrect halo model affects the
mass estimate. The datasets used to produce these histograms are
generated according to a shadow tracer profile with $\as = 100$ kpc,
but are assumed to be a power-law tracer population with $\gamma =
4.0$ when applying the Bayesian analysis. It is clear from the
comparative narrowness of the histograms in Fig.~11 (c) that the
effects of this modelling uncertainty are less serious than problems
caused by the measurement errors and the velocity anisotropy (as well
as the small size of the dataset to be discussed in the next
Section). For example, the standard deviation of the mass estimates in
the case in which proper motions are included (the solid line in
Fig.~11 (c)) is $15\%$ of $M$ and is thus much less than the intrinsic
spread due to other causes.  Only if our assumptions regarding the
satellite number density are grossly incorrect can such modelling
uncertainty be a grave problem. Serious incompleteness in the dataset
might be a cause of such blundering. However, it does seem that our
dataset can be missing at most only a few satellites.  For example,
Kleyna, Geller, Kenyon \& Kurtz (1997) argue that the current sample
is complete to the limits of current surveys. By extrapolating the
luminosity function in the absence of a cut-off, they suggest that by
surveying all the sky $\sim 1.5$ magnitudes deeper, perhaps a further
$\sim 5$ dwarfs may be recovered.

As is evident from the work in Sections~3 and~4, the decision whether
to fix the velocity normalisation or allow it to vary is another
modelling uncertainty. Fig.~11 (d) shows the spread in mass estimates
obtained from datasets of 30 points with radial velocities only, each
dataset being generated for $v_0 = 220$ \kms but being analysed with
$v_0$ as a free parameter. While the figure clearly shows a systematic
underestimate of the mass, it is important to note that this effect is
mainly due to the small size of the dataset. Comparison with the
dashed curve in Fig.~12 (a), which presents the results from
simulations in which $v_0$ is assumed to be known, shows that the
velocity normalisation uncertainty is dominated by the statistical
noise in the case of 30 data points.  We conclude that lack of
knowledge of the velocity normalisation does not degrade our current
mass estimate significantly, although it does affect our ability to
estimate $v_0$ and the scalelength $a$ as individual parameters.

\beginfigure{12}
\centerline{\psfig{figure=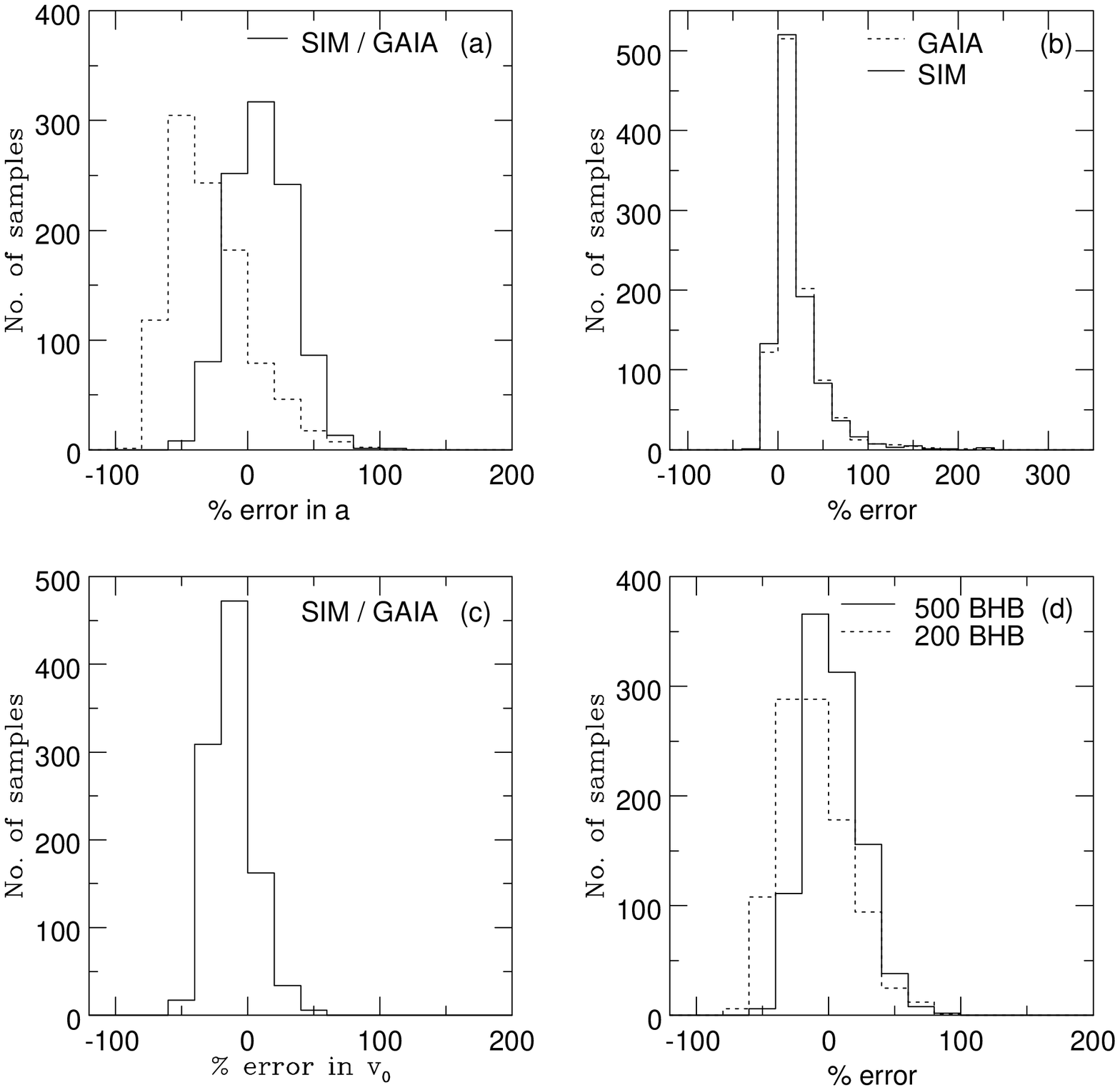,height=\hssize}}
\smallskip\noindent
\caption{{\bf Figure 12.} Histograms illustrating the impact of future
developments on the determination of $M$. As in other figures, the
histograms show how many out of 1000 datasets yielded a given
percentage error in $M$. (a) 30 points with radial velocities only
(dashed line) and with both radial velocities and proper motions
(solid line) -- as indicated, the latter case is how the dataset
should look after the SIM and GAIA missions; (b) Comparison of the
effects of measurement errors at the level of both GAIA (dashed line)
and SIM (solid line); (c) Uncertainty in the velocity normalisation
from datasets of 30 data points with radial velocities and proper
motions; (d) 200 points (dashed line) and 500 points (solid line) with
radial velocities only. Both these histograms assume a magnitude
cut-off of m$_{\rm v} < 21.5$ and take account of the spread in
intrinsic magnitudes of BHB stars. See text for discussion.}
\endfigure

\section{Future Prospects}
The mass of the Milky Way is presently fixed by a dataset of 27
objects with known distances and radial velocities, of which 6 also
possess measured proper motions. This is evidently a scanty dataset on
which to base measurements of one of the most fundamental Galactic
parameters. So, there is a pressing need for more data. What are the
prospects for the future? Here, we consider the effects of forthcoming
space missions in Section 6.1 and the new generation of large
telescopes in Section 6.2.

\subsection{The Astrometric Satellites}

As the sample of satellite galaxies is nearly complete, the dataset
can be extended only by measurements of their proper motions. Here,
the outlook is good, with the {\it Space Interferometry Mission} (SIM)
and the {\it Global Astrometry Interferometer for Astrophysics} (GAIA)
satellites scheduled to obtain microarcsecond astrometry on objects
brighter than $V= 20$. SIM is a pointing instrument and so will look
at relatively few objects with great accuracy. GAIA is a scanning
instrument with poorer accuracy but it will prove powerful for
statistical analyses of larger samples.  For SIM, wide angle
astrometry is planned to yield proper motions accurate to $\sim 2
\muay$ for $V=20$ objects, though this requires long integration times of
$\gta 4$ hours. As time on the instrument may well be at a premium,
this may mean that SIM will examine only a limited number of faint
objects and perhaps only some of the satellite companions of the Milky
Way.  The colour-magnitude diagrams of even the distant Leo~I show the
tip of the giant branch is still visible at $V=20$ (Caputo, Cassisi,
Castellani \& Marconi 1998; Hernandez-Doring, Valls-Gabaud \& Gilmore
1999).  So, even for Leo~I, SIM can find the proper motions to $\sim
5$ \kms by observing one or two stars (the internal velocity
dispersion is of course much less than the systemic proper motion of
the dwarf galaxy).

For GAIA, the target is $10 \muay$ in proper motion accuracy at $V=15$
and $100-200\muay$ at $V=20$.  The poorer accuracy of GAIA means that
the individual proper motions of bright stars at the distance of Leo~I
are still only accurate to $\sim 240$ \kms. However, GAIA will measure
the proper motions of all the stars brighter than $V= 20$, and
therefore the proper motion of the satellite can be recovered to good
accuracy, as we now show.  Caputo et al.'s (1998) colour-magnitude
diagram is derived from three {\it Wide Field Planetary Camera}
(WFPC2) frames with the aperture centered on Leo~I. The field of view
is $\sim 1.7 \sqarcmin$.  There are $\sim 50$ stars brighter than
$V=20$ visible on the colour-magnitude diagram. Leo~I subtends perhaps
$\sim 10 \sqarcmin$ on the sky, using the exponential radius given in
Mateo (1998).  In total, therefore, Leo~I has perhaps $\sim 290$ stars
brighter than $V=20$, and so the error on the proper motion of the
ensemble is less by a factor of $\sim 17$. In other words, the
components of the space motion of Leo~I are obtainable to an accuracy
of perhaps $\sim 14$ \kms with GAIA.  For closer satellites like Draco
and Ursa Minor, the situation is even more favourable.
Hernandez-Doring et al. (1999) provide a colour-magnitude diagram for
Ursa Minor which has $\sim 17$ stars brighter than $V=20$ and is
derived from single chip WFPC2 observations. Each chip represents a
field of view of $0.6
\sqarcmin$.  Taking the exponential scalelength as $8.0$ arcmin (Mateo
1998), then the number of stars in Ursa Minor brighter than $V=20$ is
$\sim 5700$.  So, GAIA can provide the components of the space motion
of Ursa Minor to an accuracy of $\sim 1$ \kms. To analyse the
implications of SIM and GAIA, it is thus reasonable to assume that
they can provide the space motions to better than $10 \%$, though
the distances of the objects may not be substantially improved.

Let us now investigate both the likely error caused by the small
number of data points available, as well as future prospects from SIM
and GAIA. We generate 1000 datasets with 30 data points, both for the
case in which knowledge of only radial velocities is assumed and that
in which the full space velocities are presumed to be measured by a
combination of SIM and GAIA.  The results are reported in
Fig.~12. From Fig.~12 (a), we see that when the number of data points
is 30, and only radial velocity data are used, the probability of
obtaining an estimate of $M$ which differs from the true value by more
than a factor of two is about $30\%$. There is also evidence for a
systematic underestimate in the mass when only radial velocities are
used. This result is true whether or not the value of $v_0$ is held
fixed during the Bayesian analysis -- this can be seen by comparing
the dashed curve in Fig.~12 (a) with Fig.~11 (d). We note that this
underestimate probably represents the worst case since the artificial
data were generated from an isotropic model ($\beta = 0$), but were
analysed assuming the uniform energy prior on $\beta$. As noted
previously, this prior is biased towards radial anisotropy. In the
present case, this bias causes the Bayesian algorithm to
systematically underestimate the kinetic energies of the satellites by
assuming that most of their motion is contained in their observed
radial velocities, which leads to a systematic underestimate of the
total mass.

When we include tangential velocities, this systematic effect is
removed and the probability of obtaining a mass estimate more than a
factor of two different from the true value is reduced to just
$2\%$. We conclude that SIM and GAIA have the potential to improve
matters substantially by removing the bias to lower masses which is
present if only radial velocities are available.

Fig.~12 (b) illustrates how the reduction of the proper motion errors
will dramatically reduce the uncertainty due to measurement errors
described in Section~5. To produce this figure, datasets of 30 data
points with both radial velocities and proper motions were
generated. Measurement errors of 100 $\muay$ and $2 \muay$ were
assumed for the GAIA and SIM missions respectively. In the case of
GAIA, it is assumed that for each dwarf galaxy typically $400$ stars
brighter than V=20 are observed, thereby reducing the error in the
individual proper motions by a factor of 20. From Fig.~12 (b), we find
that the spread in mass estimates due to measurement uncertainties for
both GAIA and SIM is $\sim 18\%$, a huge improvement on the current
situation (see Fig.~10 (a)).  SIM and GAIA also have the potential to
reduce the uncertainty in the velocity normalisation alluded to in
Section~5. Fig.~12 (c) presents a histogram of velocity normalisation
estimates from datasets of 30 data points with radial velocities and
proper motions. The peak of the histogram lies within 20$\%$ of the
true value and the mean absolute deviation is just $16\%$. Thus, after
SIM and GAIA it will certainly be possible to assess whether the
velocity normalisation of the halo is very different from $220$ \kms

Following the SIM and GAIA missions, all the major sources of error
will have been reduced to below the levels of the statistical noise
illustrated by the solid histogram of Fig.~12 (a). The average
absolute deviation caused by the sparse dataset is $\sim 20\%$ and
this represents the best that can be achieved with the satellite
galaxy dataset.

\subsection{Blue Horizontal Branch Stars}

The only option for substantially increasing the size of the dataset
is to use distant spheroid stars, especially the comparatively bright
blue horizontal branch (BHB) stars.  For example, Warren and
collaborators (1998, private communication) have begun a program of
surveying selected fields in the Milky Way halo for BHB stars and plan
to amass a dataset of $\sim 200$ with accurate distances and radial
velocities within the next few years.  Miller (1998, private
communication) reports that the 2df survey has discovered $\sim 1000$
blue horizontal branch stars in a patch of the sky covering $750$
square degrees. The present radial velocities are too crude to be of
direct use in measuring the mass of the Milky Way halo. However, the
dataset could be re-observed from the ground with the {\it Very Large
Telescope} (VLT) to provide accurate radial velocities with only a
modest investment of telescope time.  The advantage of using stellar
tracers of the distant halo is partly offset by complexity of
modelling, as the selection effects have to be taken into account. In
what follows, we modify the procedure of Section~2.3 to take account
of two factors. First, the BHB stars have a distribution of absolute
magnitudes, which we assume to be uniform in the range
[$\Mmin,\Mmax$]. By studying the dataset of Flynn et al. (1995), it
seems reasonable to take $\Mmin = 0.5$ and $\Mmax = 1.0$ for BHB
stars. Second, we can only observe stars brighter than a certain
magnitude threshold $\mt$ (which we take to be $\mt$ = 21.5, an
optimistic estimate for the VLT). This means that the probabilities
$P(r,v_r|a,\beta)$ (from Table~1) are multiplied by the selection
factor $\epsilon (s)$
\eqnam{\prob_correct}
$$\epsilon (s) = {\displaystyle \mt - 5 \logten s - 10 - \Mmin \over 
\displaystyle \Mmax - \Mmin}\quad \smin <s < \smax\eqno\new$$
Here, $s$ is the heliocentric position in kpc, and we have defined
$$\eqalign{\smin =&  10^{0.2(\mt-\Mmax -10)},\cr
\smax =& 10^{0.2(\mt-\Mmin -10)}.}\eqno\new$$
This selection factor (\prob_correct) is unity when $s < \smin$ and
vanishes when $s > \smax$.  Our procedure is to generate positions and
velocities for the BHB stars from a power-law tracer whose density
falls off like $\sim r^{-3.5}$. We then choose an absolute magnitude
uniformly from our assumed uniform distribution of intrinsic
magnitudes and test to see if this BHB star lies in the observable
sample.  In this way, simulated datasets of 200 and 500 BHB stars are
generated and then analysed with the Bayesian algorithm, incorporating
of course our new selection factor (\prob_correct).

Fig.~12 (d) shows histograms for 1000 datasets of samples of 200 and
500 BHB stars. In both these histograms, it is clear that the
systematic underestimate evident in Fig.~12 (a) is not
present. Samples of such sizes are large enough to evade this awkward
effect. However, a price is paid for using magnitude-limited samples,
in that the histograms have somewhat larger spreads than the
equivalent histograms for simulated data of complete
samples. Nonetheless, with a dataset of 200 BHB radial velocities, the
average absolute deviation about the mean mass estimate is just $21\%$
and with 500 points it is $17\%$. These numbers clearly illustrate the
value of using the BHB stars to augment the satellite dataset.  BHB
datasets will remove any possible problem with the systematic
underestimate that is present in the much smaller satellite galaxy
dataset.  Measurement errors are less important than the statistical
effect that comes from the bias in the sample introduced by the
selection factor. It is this that causes the broadening of the dashed
histogram in Fig.~12 (d). Assembling a dataset of $\sim 200$ BHB stars
with radial velocities and distances accurate to $\sim 10 \%$ is very
clearly worth doing, as the investment of telescope time is not
substantial compared to the scientific pay-off. The advantages of
extending the BHB dataset to $\sim 500$ stars appear to be slight --
the average absolute error is reduced by only $\sim 4 \%$, although
the peak in the histogram is more centred on the true value.

\section{Conclusions}

\beginfigure{13}
\centerline{\psfig{figure=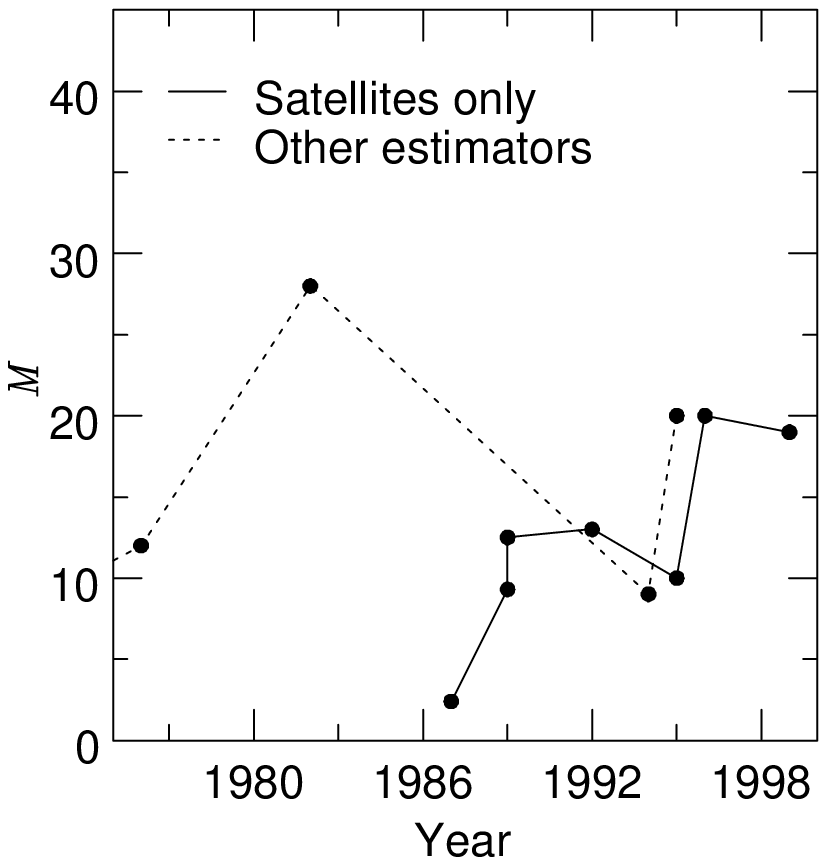,width=\hssize}}
\smallskip\noindent
\caption{{\bf Figure 13.} Recent estimates of the total Mass of the
Milky Way (in units of $10^{11} M_{\odot}$). Those based on satellite
motions only are shown by the solid line, while those based on other
arguments (e.g. Local Group timing) are shown by the dotted
line. Sources: Einasto, Haud, Joeveer \& Kaasik (1976), Lin \&
Lynden-Bell (1982), Little \& Tremaine (1987), Zaritsky, Olszewski,
Schommer, Peterson \& Aaronson (1989), Kulessa \& Lynden-Bell (1992),
Byrd, Valtonen, McCall \& Innanen (1994), Lin, Jones \& Klemola
(1995), Peebles (1995), Kochanek (1996), this paper (1999)}
\endfigure

\noindent
This paper has presented a consistent estimate of the mass of the
Milky Way. Previous analyses have given different answers depending on
whether or not Leo~I is included in the dataset. Our modelling has
advanced the debate by providing a consistent answer irrespective of
the presence or absence of Leo~I, provided both the radial velocity
and available proper motions are used. The consequent mass estimates
and likelihood contours are in much better agreement than previously
obtained. This happy circumstance is caused partly by the improved
information on the proper motions of the dwarfs and partly by the new
halo model.  By generating simulated data, it is also straightforward
to answer the question: How likely is it that, in a dataset of $\sim
30$ satellites with known radial velocities, there is an object like
Leo~I whose inclusion or exclusion changes the inferred mass in a
dramatic way (or, more specifically, by a factor of $\sim 5$)~?  This
is actually not common, happening only some $\sim 0.5 \%$ of the
time. Although prior expectation does not favour the existence of a
Leo~I, such a happenstance is not impossible (the probability is
small, but not zero).

Our best estimate is a total mass of the Milky Way halo of $\sim 1.9
\times 10^{12} M_{\odot}$ and a halo scalelength of $\sim 170$
kpc. What is the likely error in this mass estimate?  Using synthetic
datasets of radial velocities of $30$ satellites, we have shown that
there is a systematic tendency to underestimate the mass. The
probability of obtaining a mass estimate which is smaller than the
true value by more than a factor of two is $\sim 30 \%$.  From our
analysis of the likely sources of error in Section~5, we conclude that
-- in addition to the systematic effect caused by the small size of
the dataset -- measurement errors are the most troublesome with the
uncertainties in the proper motions being the main culprits. The net
effect of these two main sources of error is a spread with a
half-width of $\sim 90 \%$ coupled with a possible systematic
underestimate of a factor of two.  Taking this into account, our value
for the mass of the Milky Way halo including the errors is $\sim
1.9^{+4.0}_{-1.9} \times 10^{12} M_{\odot}$.  Fig.~13 shows a graph of
recent mass estimates of the Milky Way halo based solely on the
motions of the satellites and globular clusters (solid line), together
with those based on other arguments (dotted line).  Over the past 15
years, there has been a tendency for an increase in the mass estimates
obtained from satellite radial velocities due to the increased size of
the dataset and the availability of more proper motions.  Our mass
estimate is a slight decrease on the most recent previous
determination by Kochanek (1996) and but it still fits into this
general trend.  We note that the mass estimates obtained from a
variety of methods are now in good agreement. Zaritsky (1998) makes
the point that the data from a variety of sources are consistent with
an isothermal sphere of amplitude 180 - 220 \kms extending outwards to
$\gta 200$ kpc, a conclusion which agrees well with our results.

We have also explored the effects of allowing the normalisation $v_0$
to vary as a free parameter. When the prior probability is $1/v_0^2$
we obtain a most likely value of $\sim 280$ \kms, independent of the
presence or absence of Leo~I. The most likely values of $a$ are $110$
kpc when Leo~I is included and $50$ kpc when Leo~I is excluded,
leading to mass estimates of $2.0 \times 10^{12} M_{\odot}$ and $0.9
\times 10^{12} M_{\odot}$ respectively. This analysis also yields
strongly tangential values for the velocity anisotropy $\beta$.
However, all the results of the 3-parameter fitting exhibit a strong
sensitivity to the choice of prior probabilities for $v_0$ and $\beta$
which suggests that they should not be given too much weight. We
conclude that at present the small amount of data, crucially in the
area of satellite proper motions, means that it is not feasible to
constrain $v_0$ with any degree of confidence.

The mass of the Milky Way halo within 50 kpc is $\sim
5.4^{+0.2}_{-3.6} \times 10^{11} M_{\odot}$.  This is a more robust
quantity than the total mass. Our error estimates are inferred from
the maximum and minimum values of the total mass. Note that the errors
on the mass within 50 kpc are asymmetrically distributed about the
most likely value in the opposite sense to the errors in the total
mass.  This seems slightly counterintuitive. The reason is that
increasing the total mass of the halo above $\sim 1.9 \times 10^{12}
M_{\odot}$ has little effect on the mass within 50 kpc, whereas
diminishing the total mass can cause more significant changes.  It is,
of course, the mass within 50 kpc that is the relevant figure to bear
in mind when considering interpretations of the microlensing
experiments. For example, Honma \& Kan-ya (1998) have argued that the
Milky Way need not have a flat rotation curve out to $50$ kpc and
hence suggest that the timescales of the lensing events are consistent
with brown dwarfs. The total mass in their Plummer model of the halo
is just $1.1 \times 10^{11} M_\odot$. Though this may be consistent
with the gas rotation curve (which cannot be traced beyond 20 kpc), it
is quite incompatible with the mass estimates derived from the
satellite galaxies (as well as the orbit of the Magellanic
Stream). The origin of the microlensing events towards the Large
Magellanic Cloud is presently unknown and a number of intriguing
suggestions have been made. For example, they may lie in the Large
Magellanic Cloud itself (Sahu 1994), or in an intervening stellar
population or tidal shroud (Zaritsky \& Lin 1997; Zhao 1998) or even
in the warped and flaring Milky Way disk (Evans, Gyuk, Turner
\& Binney 1998). Nonetheless, Alcock et al. (1997) assert that the 
lenses largely lie in the Milky Way halo and provide a
model-independent estimate of the halo mass in the lensing population
within 50 kpc of $2.0^{+1.2}_{-0.7} \times 10^{11} M_\odot$. If their
assumption as to the location of the lenses is correct, the
microlensing results imply that $\sim 36 \%$ of the halo within 50 kpc
is baryonic with the remainder of the halo being built from elementary
particles or baryonic objects (such as cold molecular clouds) that do
not produce microlensing. However, as can be deduced from the error
bounds, the baryonic fraction is not well constrained at the moment.

Current estimates of the total mass of the Local Group set it at $\sim
4-8 \times 10^{12} M_{\odot}$ (e.g., Peebles 1996, Schmoldt \& Saha
1998). Based on their asymptotic rotation curves, M31 is $\sim 30\%$
more massive than the Milky Way.  Given our result for the Milky Way
halo, this implies that the mass of M31 is $\sim 3.0 \times 10^{12}
M_{\odot}$. We conclude that more than $50\%$ (and perhaps almost all)
of the mass in the Local Group is concentrated around the two largest
group members. Let us note that these results receive confirmation
from recent work of Peebles using his numerical action method. For
example, Peebles (1995, 1996) obtains a mass for the Milky Way halo of
$\sim 2 \times 10^{12} M_\odot$ and a mass for M31 of $\sim 3.4 \times
10^{12} M_\odot$ using the motions of the most distant Local Group
satellites.

The coming decade promises to be fruitful in terms of the availability
of new data.  We have shown that obtaining radial velocities for large
samples of blue horizontal branch stars can provide a very promising
line of attack on the problem. A dataset of even 200 such stars will
reduce the statistical uncertainty in the mass estimate to $\sim 21
\%$, as well as removing the possible systematic effect that occurs
with the small samples of satellite galaxies. This illustrates that
the scientific returns from such a program could be high for a
relatively low investment of telescope time.  In the longer term, SIM
and GAIA will be able to measure the proper motions of all the Milky
Way satellites. For example, using the colour-magnitude diagrams, we
have shown that the proper motions of the most distant dwarfs like
Leo~I will be determined to $\sim 5-14$ \kms, while the nearer dwarfs
like Ursa Minor will be determined to $\sim 1$\kms. This will provide
the mass of the Milky Way halo to within $\sim 20\%$ and will also
allow the amplitude of the velocity normalisation to be determined to
within $\sim 16\%$.

It has been suggested by Johnston et al. (1998) that SIM and GAIA may
be used to measure the proper motions of stars in a stream and hence
to find the mass of the Milky Way halo. In particular, they suggest
that measuring the proper motions of 100 stars brighter than 20th
magnitude in a tidal stream to $\sim 4 \muay$ may be enough to
determine the mass of the Milky Way to a few percent. Such an accuracy
on the proper motions is not achievable at $V = 20$ for GAIA. For SIM,
wide angle astrometry to this accuracy requires long integration times
of $\gta 4$ hours. The mass of the Milky Way within 50 kpc is
reasonably certain, and it is data collected on objects beyond 50 kpc
(and preferably beyond 100 kpc) that is most helpful in discriminating
between models (see, for example, Fig.~A1 of Lynden-Bell \&
Lynden-Bell 1995).  Unfortunately, there is no known stellar stream at
such large Galactocentric radii for use as a SIM target.  At large
distances, the proper motion errors of individual stars will remain
large with available technology, thus frustrating any certain
identification of stream candidates.  It is also important to assess
the effects of some of the assumptions made by Johnston et al. (1998)
before the figure of a few percent error can be accepted, as it does
not include all the modelling and measurement errors.

As data beyond 50 kpc is scarce, we believe that the optimal approach
is to use every scrap of information. We believe that the future will
belong to joint analyses of the datasets of both the radial and proper
motions of the satellites together with large samples of distant BHB
stars. This is the best strategy for reducing both the statistical
noise and the measurement uncertainties.

\section*{Acknowledgments}
NWE thanks the Royal Society for financial support, while MIW
acknowledges help from a Scatcherd Scholarship. We thank Geza Gyuk,
Xavier Hernandez-Doring, Paul Hewett, Eamonn Kerins, Mike Irwin, Jenny
Read, David Valls-Gabaud and Steve Warren for helpful discussions and
suggestions. Gerry Gilmore, Michael Perryman and Tim de Zeeuw provided
useful information on the GAIA satellite. We also thank the referee,
Scott Tremaine, for several insightful comments on this work.

%
%

\eqnumber =1
\def\chaphead{\hbox{A}}
\appendix
\section{A Family of Error Convolution Functions}
In this appendix, we present the properties of a family of error
convolution functions which we call the generalised Lorentzian
family. These functions are already known in the plasma physics
literature (Vasyliunas 1968, Summers \& Thorne 1991), but do not seem
to be readily available in the astronomical literature. The $n$th
member of this family $E_{n}$ is given by
$$E_{n}(v) = {1 \over \sqrt{2\pi} n\sigma_{L,n}}{\Gamma[n] \over
\Gamma[n-1/2]} {(2n\sigma_{n}^2)^n \over
(2n\sigma_{n}^2+v^2)^n}.\eqno\new$$
The $n=1$ member is the Lorentzian $E_1$ which we use in our
calculations to take account of the observational errors in the proper
motions of the satellites. In the limit $n\rightarrow\infty$, the
error convolution function becomes
$$E_{\infty}(v) = {1 \over \sqrt{2\pi}\sigma} e^{v^2 \over
2\sigma^2},\eqno\new$$
which is the familiar Gaussian. The convergence to a Gaussian is very
rapid with increasing $n$ and, as Fig.~A1 shows, $E_{3}$ is already a
close approximation to a Gaussian, although it remains somewhat
broader.

To normalise the $E_{n}$ family, we demand that the quartiles for
each member be the same as those of a Gaussian of width
$\sigmaG$. For a Gaussian, the quartiles $\xG$ are given by
solving
$${\rm Erf}\,\Bigl[{\xG\over\sqrt{2}\sigmaG} \Bigr] = 
{2 \over \sqrt{\pi}}\int_0^{\xG/(\sqrt{2}\sigmaG)} e^{-t^2}\,dt
={1\over 2}.\eqno\new$$
Hence, we find that $\xG = 0.67449\sigmaG$. To find the corresponding
value of $\sigma_{n}$, we must solve the integral equation
$${1 \over \sqrt{2\pi} n\sigma_{L,n}}{\Gamma[n] \over
\Gamma[n-1/2]} \int_{-\xG}^{\xG}\,dv {(2n\sigma_{n}^2)^n \over
(2n\sigma_{n}^2+v^2)^n} = {1 \over 2},\eqno\new$$
to obtain $\sigma_{n}$ in terms of $\xG$ (and hence in terms of
$\sigmaG$). This is analytically tractable only for $n=1$, when we
obtain $\sigma_{1} = 0.4769\sigmaG$. For all other values of $n$,
$\sigma_{n}$ must be found numerically. Some numerical values are
presented for convenience in Table~B1.

\bye